\documentstyle[aps,twocolumn,epsf]{revtex}

\begin{document}

\begin{center}
{\bf Canonical Ensemble of} \\ 
\vspace{0.2cm}
{\bf Initial States Leading to}\\ 
\vspace{0.2cm}
{\bf Chiral Fluctuations}

\vspace{0.7cm}
{\sc D\'enes Moln\'ar}
\\
\vspace{0.3cm}
Physics Department, University of Bergen \\
N-5007 Bergen, All\'egaten 55, and\\
Physics Department, Columbia University\\
New York, NY 10027, USA\\
\vspace{0.7cm}
{\sc L\'aszl\'o P. Csernai}
\\
\vspace{0.2cm}
Physics Department, University of Bergen \\
N-5007 Bergen, All\'egaten 55, Norway\\

\vspace{0.2cm}
and \\

\vspace{0.7cm}
{\sc Zsolt I. L\'az\'ar}
\\
\vspace{0.2cm}
Physics Department, University of Bergen \\
N-5007 Bergen, All\'egaten 55, Norway\\

% revision number for authors
%{\bf Sixth revised version June 24, 1998 \\}  

% revision counter for (re)submission
%{\bf Initial submission Nov 24, 1997 \\}  
{\bf Final revision June 24, 1998 \\}  

\end{center}

\vspace{0.5cm}
\centerline {\bf Abstract}
\vspace{0.5cm}

\renewcommand{\baselinestretch}{1.0}

\footnotesize 
In energetic heavy ion collisions, if quark-gluon plasma is formed, its
hadronization may lead to observable critical fluctuations, i.e., DCC
formation. The strength and observability of these fluctuations depend on
the initial state. Here we study the canonical ensemble of initial
states of chiral fluctuations in heavy ion collisions and the probability to
obtain observable domains of chiral condensates.

\vspace {0.4cm} 
PACS number(s): 12.39.Fe, 12.38.Mh, 25.75.-q

\renewcommand{\baselinestretch}{1.0}

\normalsize

\vspace{0.7cm}
\noindent

\section{\bf Introduction}

\vspace{0.2cm}
In a recent article the initial states of Disoriented Chiral Condensate 
(DCC) were studied, the importance of non-vanishing chiral 'angular 
momentum' was pointed out, and the microcanonical ensemble of initial 
states were evaluated for fixed energy and chiral angular momentum
\cite{birmol97}.  These results indicated that the observability
of DCC formation is a delicate problem, and it strongly depends on the 
reaction mechanism of heavy ion collisions.

Since at this moment reaction models which would predict a chiral angle
distribution at the onset of hadronization do not exist, we assume that
thermal equilibrium is reached in the supercooled plasma at the freeze-out,
$T_{FO}=T_0<T_c$, which coincides with the beginning of the process of
hadronization\cite{csocse94,csemis95}.
Consequently we also assume that at this moment the chiral degrees of
freedom in the system are thermalized and thus can be characterized by
a canonical distribution corresponding to the freeze-out temperature,
$T_0$, and to the volume of the system, $V$. (This latter quantity is
vital
because we study fluctuations in a small, finite system!)  
Unless further constraints on the chiral degrees of freedom can be 
established from more detailed reaction models the generated canonical
distribution of chiral initial states will represent the statistical
distribution in heavy ion experiments.

Having obtained this distribution we will analyze which initial 
configurations lead to strongly fluctuating domains of DCC, and what are
the overall chances to detect these experimentally.

For our studies, just like ref. \cite{birmol97},
we have chosen the linear sigma
model because it contains
the relevant chiral degrees of freedom and it is still relatively simple.

\section{The linear sigma model}
\label{Sc:Sigma}

The linear sigma model involves the isodoublet of the light quark
fields, $\Psi$, and a real, four-component chiral
field,
$\Phi\equiv\left(\sigma,\vec\pi\right)$, where $\sigma$ is a scalar
field,
while $\vec\pi\equiv\left(\pi_1,\pi_2,\pi_3\right)$ is a triplet of
pseudoscalar pions. The Lagrangian density is
\footnote{We shall use $\hbar=c=k=1$ units.}
\begin{eqnarray}
  {\cal L}&=&\bar\Psi\left[i\gamma_\mu\partial^\mu
              -g\left(\sigma+i\gamma_5\vec\tau\vec\pi\right)\right]\Psi
              \nonumber\\
          &+&\frac{1}{2}\left(\partial_\mu\sigma\right)
                      \left(\partial^\mu\sigma\right)
          +\frac{1}{2}\left(\partial_\mu\vec\pi\right)
                      \left(\partial^\mu\vec\pi\right)
           - U\left(\sigma,\vec\pi\right) \ ,
   \label{Lsigma}
\end{eqnarray}
with the so-called "Mexican hat" potential energy density
\begin{equation}
    U\left(\sigma,\vec\pi\right)= 
    \frac{\lambda}{4}
       \left(\sigma^2+\vec\pi^2-v^2+\frac{T^2}{2}\right)^2
          %-H\sigma \ .
          -H\sigma + U_0(T)\ ,
    \label{MHpot}
\end{equation}
where the last term, $U_0(T)$, ensures that
the {\em absolute} minimum of the potential energy density is 0.
Without the
$H\sigma$ term, this Lagrangian density is invariant under the chiral
$SU_L(2)\otimes SU_R(2)$ transformations. 

The parameters in this Lagrangian are chosen in such a way \cite{csemis95} 
that in {\em normal vacuum} at $T=0$ chiral symmetry is spontaneously
 broken and the expectation values of the meson fields are
\begin{equation} \label{nvac}
   \langle \sigma \rangle=f_{\pi},~~\langle{\vec{\pi}}\rangle=0\ , 
\end{equation}
where $f_{\pi}$=93 MeV is the pion decay constant.  In the vacuum with
broken chiral symmetry pions represent soft ``azimuthal'' excitations of
the chiral field.  To have the correct pion mass in vacuum ($T=0$),
$m_{\pi}$=138
MeV, one should take
\begin{equation} 
   v^2=f_{\pi}^2-\frac{m_{\pi}^2}{\lambda},\ \ \ \ \ H=f_{\pi}m_{\pi}^2\ .
   \label{par}
\end{equation}
The parameter $\lambda$ is related to the sigma mass, 
$m_{\sigma}^2 = 2\lambda f_{\pi}^2+m_{\pi}^2$, which can be chosen to be
about 600MeV (then $\lambda \approx 20$).  Sigmas represent stiff,
``radial'' excitations of the chiral field.  The remaining coupling
constant $g$ can be fixed by the requirement that the effective quark
mass, $m_q^2 = g^2 (\sigma^2 + \vec{\pi}\/^2)$, 
in broken vacuum
%, $m_q=g\langle \sigma \rangle=gf_{\pi}$,
coincides with the constituent quark mass in hadrons, which is about 1/3
of the
nucleon mass, $m_N$.  This gives $g\approx (m_N/3)/f_{\pi}=3.36$.

\section{Ideal Hydrodynamics in the Local Rest Frame}
\label{Ch:Exp}

When solving the equations of motion corresponding to our Lagrangian
(\ref{Lsigma}) 
we shall assume a one-dimensional Bjorken scenario \cite{bjo83}.
Consequently,
we shall only consider field evolutions having one-dimensional scaling
property and we shall work 
in the corresponding local rest (LR) frame. Therefore, let us 
briefly review here the properties of a scaling hydrodynamical expansion
in the LR frame.

\subsection{Framework: the Bjorken model}

\label{Sect:LR}

The local rest frame is a curvilinear coordinate system where the matter
is at rest. 
%%%%%%%%%%%%%%%%%%%%%%
% initial submission %
%It is easy to verify that for a longitudinal scaling expansion
%%%%%%%%%%%%%%%%%%%%%%
% 1st revision       %
For a longitudinal scaling expansion
%%%%%%%%%%%%%%%%%%%%%%
along the $z$-axis the LR frame coordinates, $x^i\equiv(\tau,x,y,\eta)$,
 are given by the c.m. coordinates, $x^\mu\equiv(t,x,y,z)$, as 
\begin{equation}
  \tau\equiv\sqrt{t^2-z^2} \qquad \hbox{ \rm and  } \qquad 
  \eta\equiv\frac{1}{2}\ln\frac{t+z}{t-z}.  
  \label{deftaueta} 
\end{equation}
Thus, the invariant space-time interval in the LR frame is
%%%%%%%%%%%%%%%%%%%%%%
% initial submission %
%$$
% ds^2\equiv dt^2-dx^2-dy^2-dz^2=d\tau^2-dx^2-dy^2-\tau^2d\eta^2,
%$$
%%%%%%%%%%%%%%%%%%%%%%
% 1st revision       %
$ds=d\tau^2-dx^2-dy^2-\tau^2d\eta^2$,
%%%%%%%%%%%%%%%%%%%%%%
therefore the covariant metric tensor is
\begin{equation}
   g_{ik}={\rm diag} (1,-1,-1,-\tau^2).
   \label{metrictensor}
\end{equation}
This metric tensor gives rise to three non-vanishing Christoffel symbols:
%%%%%%%%%%%%%%%%%%%%%%
% initial submission %
%$$
%  \Gamma^{0}_{33}=\tau, \qquad
%  \Gamma^{3}_{30}=\Gamma^{3}_{03}=\frac{1}{\tau}.
%$$
%%%%%%%%%%%%%%%%%%%%%%
% 1st revision       %
$\Gamma^{0}_{33}=\tau$, $\ \Gamma^{3}_{30}=\Gamma^{3}_{03}=1/\tau$.

%%%%%%%%%%%%%%%%%%%%%%
The equations of ideal hydrodynamics 
are the conservation laws given by the
requirement that the energy
momentum tensor, $T^k_i$, has a vanishing divergence, i.e.,
$T^k_{i;k}=0$.\cite{LanLif2}
%%%%%%%%%%%%%%%%%%%%%%
% initial submission %
%This divergence can be expressed in terms of the partial derivatives
%and the Christoffel symbols; 
%after a short calculation
%%%%%%%%%%%%%%%%%%%%%%
% 1st revision       %
Expressing this divergence in terms of the partial derivatives
and the Christoffel symbols, 
%%%%%%%%%%%%%%%%%%%%%%
we arrive to the following four equations:
\begin{eqnarray}
  T^k_{0;k}&=&\frac{\partial{T^k_0}}{\partial x^k}
       +\frac{1}{\tau}\left(T^0_0-T^3_3\right) \ , \qquad 
  T^k_{1;k}=\frac{\partial{T^k_1}}{\partial x^k}
       +\frac{1}{\tau} T^0_1 \ , \nonumber\\
  T^k_{2;k}&=&\frac{\partial {T^k_2}}{\partial x^k}
       +\frac{1}{\tau} T^0_2 \ ,
  \qquad\qquad\quad
  T^k_{3;k}=\frac{\partial{T^k_3}}{\partial x^k}
       -\tau T^3_0 \ .
  \label{tensordiv}
\end{eqnarray}

In local thermal equilibrium, if we neglect the effects of viscosity and
heat conductivity, the energy momentum tensor has the form 
\cite{Csernai94}
%%%%%%%%%%%%%%%%%%%%%%
% initial submission %
%$$
%  T_{ik}=\left(e+p\right)u_i u_k-p\;g_{ik},
%$$
%%%%%%%%%%%%%%%%%%%%%%
% 1st revision       %
$T_{ik}=\left(e+p\right)u_i u_k-p\;g_{ik}$,
%%%%%%%%%%%%%%%%%%%%%%
where $e$ is the energy density, $p$ is the pressure in the co-moving
frame, and $u_i$ is the flow
velocity of the matter. Thus, in the LR frame the energy momentum tensor
is diagonal, i.e., 
\begin{equation}
  T^k_i={\rm diag}\left(e,-p,-p,-p\right).
  \label{diagenmomtens}
\end{equation} 
The conservation laws
(\ref{tensordiv}) lead to
\begin{equation}
    \frac{\partial{e}}{\partial\tau}+\frac{p+e}{\tau}=0, \qquad
    \frac{\partial p}{\partial x}=\frac{\partial p}{\partial y}
    =\frac{\partial p}{\partial\eta}=0.
    \label{conslaws}
\end{equation}
These, together with the charge conservations
$$ 
  D_i \left(\rho_c u^i\right)=0 ,
$$
(where $c$ runs for all conserved charges, and
$D_i$ stands for the covariant derivative)
and the equation of state (EoS) which
connects the pressure, energy density, and the {\em proper} charge
densities
$\rho_c$ [$\rho_c \equiv dN_c /(dx\, dy\, d\eta)$],
form a closed set of equations, which can be solved for given initial
and boundary conditions. 
In the LR frame the charge conservations simplify to
\begin{equation}
   \frac{\partial   \rho_c}{\partial \tau}=0 
  \label{chargecons}.
\end{equation}

The equation of the Bjorken model
 can be
obtained by imposing 
%%%%%%%%%%%%%%%%%%%%%%
% initial submission %
%
%boost invariance and transverse translational invariance, i.e.,
%that all quantities, $e,\ p$ and $\rho_c$, should be
%independent of $\eta$, $x$ and $y$. 
%This implies that 
%%%%%%%%%%%%%%%%%%%%%%
% 1st revision       %
the boost invariance and transverse translational invariance 
of $e,\ p$ and $\rho_c$.
Then, by Eqs. (\ref{chargecons}),
%%%%%%%%%%%%%%%%%%%%%%
the charges are constants of motion: $\rho_c(\tau)=\rho_c(\tau_0)$.
%%%%%%%%%%%%%%%%%%%%%%
% initial submission %
%From Eqs. (\ref{conslaws}) only one remains
%non-trivial and we arrive to
%%%%%%%%%%%%%%%%%%%%%%
% 1st revision       %
Eqs. (\ref{conslaws}) simplify to
%%%%%%%%%%%%%%%%%%%%%%
an initial value problem for an ordinary
differential equation in $\tau$:
\begin{equation}
    \frac{d e(\tau)}{d \tau}+\frac{p(\tau)+e(\tau)}{\tau}=0,
   \label{eqBjorken}
\end{equation}
with
%%%%%%%%%%%%%%%%%%%%%%
% initial submission %
%$$ e(\tau_0)=e_0,\ \ p(\tau_0)=p_0, 
%    \ \ \rho_c(\tau_0)=\rho_{c0},
%$$
%%%%%%%%%%%%%%%%%%%%%%
% 1st revision       %
$e(\tau_0)=e_0$,
%%%%%%%%%%%%%%%%%%%%%%
where $p=p(e,\rho_{c0})$ is given by the EoS. 

%%%%%%%%%%%%%%%%%%%%%%%%%%%%%%%%%%%%%%%%%%%%%%%%
% initial submission - deleted in 1st revision %
%\subsection{Thermodynamics of the longitudinal scaling expansion.}
%
%The surface element of the $\tau=const.$ hyperplane is $d\Sigma=\tau
%d\eta\,
%dx\, dy$. Therefore, the volume of our longitudinally expanding system is
%proportional to $\tau$,
%i.e., $dV/d\tau=V/\tau$.
%
%With the help of the conservation laws (\ref{eqBjorken}), 
%the time derivative of the total energy, $E\equiv Ve$, in the LR frame
%can be expressed as
%\begin{equation}
%  \frac{d E}{d \tau}=V\frac{d e}{d \tau}+e\frac{d V}{d\tau}
%   =-p\,\frac{dV}{d\tau}.
%  \label{totEderiv}
%\end{equation}
%Hence, during an infinitesimal time interval $d\tau$ the 
%total energy changes by $dE=-p\,dV$,
%thus the {\em total} change of the energy corresponds to
%the work by the pressure.
%Making a comparison with the well-known
%thermodynamical
%relation \cite{Reif65}, 
%$dE=T\,dS-P\,dV$,
%we can conclude that the entropy change, $dS$, is zero during the
%expansion, thus 
%the expansion is isentropic (adiabatic).
%%%%%%%%%%%%%%%%%%%%%%%%%%%%%%%%%%%%%%%%%%%%%%%%

\subsection{Scaling expansion in the linear sigma model}

Let us consider the sigma model without quarks ($g=0$). Then 
the Lagrangian density (\ref{Lsigma}), in
a
general curvilinear coordinate system, is
\begin{equation}
  {\cal L}=\frac{1}{2}\left(\partial_i \Phi_r \right)g^{ik}
                      \left(\partial_k \Phi_r\right) 
                     - U(\Phi_r,T),
  \label{lagdenssigma}
\end{equation}
where $\partial_i \equiv\frac{\partial}{\partial x^i}$,
$\Phi_r=(\sigma,\vec\pi)$, there 
is an implied summation for indices occurring twice ($i,\ k$ and $r$),
and 
$$
  U(\sigma,\vec\pi,T)=
\frac{\lambda}{4}\left(\sigma^2+\vec\pi^2-v^2+\frac{1}{2}T^2(x^i)\right)^2
 -H\sigma+U_0,
$$
where $x^i$ is a space-time coordinate $x^i=(\tau,x,y,\eta)$.

The energy-momentum tensor, $T_{ik}$, 
%%%%%%%%%%%%%%%%%%%%%%
% initial submission %
%can be calculated 
%from the Lagrangian density, $\cal L$, 
%%%%%%%%%%%%%%%%%%%%%%
% 1st revision       %
can be obtained from the Lagrangian density 
%%%%%%%%%%%%%%%%%%%%%%
using the relation\footnote{Our sign convention in the
space-time interval is different
from that of ref. \cite{LanLif2}; therefore, we have a {\em minus}
sign on the left hand side.} \cite{LanLif2}
$$
  -\frac{1}{2}\sqrt{-g}\,T_{ik}=
     \frac{\partial}{\partial x^l}\frac{\partial \sqrt{-g}\,{\cal
L}}{\partial \frac{\partial g^{ik}}{\partial x^l}}
    -\frac{\partial \sqrt{-g}\,{\cal L}}{\partial g^{ik}},
$$
where $g$ stands for the determinant of the metric tensor, $g_{ik}$.
A straightforward calculation gives
\begin{equation}
  T_{ik}=\left(\partial_i \Phi_r \right)\left(\partial_k \Phi_r \right)
          -{\cal L}\,g_{ik}.
  \label{enmomtens}
\end{equation}
This is, unlike the energy-momentum tensor
in the Bjorken model, generally not diagonal. However, 
if we impose boost invariance, the off-diagonal elements in the
energy-momentum tensor vanish. From Eqs. (\ref{enmomtens}),
(\ref{lagdenssigma}) and (\ref{metrictensor}), we can easily obtain the
diagonal elements:
\begin{eqnarray}
  T^0_0&=&\frac{1}{2}\left(\partial_\tau \Phi_r \right)
                     \left(\partial_\tau \Phi_r \right)
         +U(\Phi_r)\equiv e_{kin}+e_{pot}\equiv e \nonumber\\
  T^1_1&=&\frac{1}{2}\left(\partial_\tau \Phi_r \right)
                     \left(\partial_\tau \Phi_r \right)
         -U(\Phi_r) 
  \equiv e_{kin}-e_{pot}\equiv -p \nonumber\\
  T^2_2&=&T^3_3=T^1_1,
\end{eqnarray}
where $e_{kin}\equiv\frac{1}{2}\left(\partial_\tau \Phi_r\right)
                     \left(\partial_\tau \Phi_r \right)$ is
the kinetic
energy density and $e_{pot}\equiv U$ is the potential energy density, and
we introduced the total energy density, $e \equiv e_{kin}+e_{pot}$, and 
$p\equiv e_{pot}-e_{kin}$.
These are
functions of the proper time $\tau$ as the fields follow the equations of
motion (EoM).

Obviously, the energy momentum tensor now has the form
(\ref{diagenmomtens}),
therefore
the conservation laws (\ref{conslaws}) and all their consequences are
also valid now. 
Particularly, this leads to
\begin{equation}
  \frac{d e}{d \tau}=-\frac{e+p}{\tau}=-2\frac{e_{kin}}{\tau} \ ,
  \label{endenssigma}
\end{equation}
which can be seen as a generalization of the usual energy
conservation for our expanding system.
%%%%%%%%%%%%%%%%%%%%%%
% initial submission %
%This equation can be used, and was used by us, 
%%%%%%%%%%%%%%%%%%%%%%
% 1st revision       %
We used this equation
%%%%%%%%%%%%%%%%%%%%%%
as a check of the numerical solution of the equations of
motion.

%%%%%%%%%%%%%%%%%%%%%%
% initial submission %
%For the total energy,
%$E\equiv V\,e$, from Eq. (\ref{totEderiv}) we have \begin{equation}
%  \frac{d E}{d \tau}=\frac{E_{pot}-E_{kin}}{\tau} \ ,
%  \label{totensigma} 
%\end{equation}
%%%%%%%%%%%%%%%%%%%%%%
% 1st revision       %
To obtain how the {\em total} energy evolves, 
we have to take the expansion into account.
The surface element of the $\tau=const.$ hyperplane 
is $d\Sigma=\tau d\eta\, dx\, dy$. 
Therefore, 
the volume of our longitudinally expanding system 
is proportional to $\tau$, i.e., $dV/d\tau=V/\tau$.
Now, 
with the help of the conservation laws (\ref{eqBjorken}), 
the time derivative of the total energy, $E\equiv Ve$, 
can be expressed as
\begin{equation}
  \frac{d E}{d \tau}= \frac{E_{pot}-E_{kin}}{\tau}
   =-p\,\frac{dV}{d\tau} \ ,
  \label{totensigma}
\end{equation}
%%%%%%%%%%%%%%%%%%%%%%
where  $E_{kin}\equiv V\,e_{kin}$ is the kinetic energy and  
$E_{pot}\equiv V\,e_{pot}$ is the potential energy of our expanding
system.
The interpretation of Eq. (\ref{totensigma}) could be similar to that of
the corresponding equation for the Bjorken model, Eq.
(\ref{eqBjorken}). However, for the
sigma model we can easily have a negative pressure, which corresponds to
an {\em increase} of the total energy during expansion! This is the case
if $E_{pot}>E_{kin}$ in Eq. (\ref{totensigma}), 
which is typical for the early evolution in the
quench scenario.
%%%%%%%%%%%%%%%%%%%%%%%%%%%
% 1st revision - inserted %
We also note that, 
as pointed out already in ref. \cite{bjo83},  
Eq. (\ref{totensigma}) means 
that we have an isentropic (adiabatic) expansion.
%%%%%%%%%%%%%%%%%%%%%%%%%%%

\section{Equations of motion}

In the mean-field approximation, ignoring all-loop contributions and
considering $\sigma$ and $\vec\pi$ as classical fields, 
we can obtain the following equations motion \cite{kapsri94,gavmul94,csemis95}
:
\begin{eqnarray}
  \partial_\mu \partial^\mu \sigma(x) + \lambda \left[\sigma^2(x){+}
  {\vec{\pi}}^2(x) {-} v'^2(x)\right] \sigma(x)&& \nonumber\\
  -H&=&\,- g \rho_S(x),
  \nonumber \\
  \partial_\mu \partial^\mu\vec{\pi}(x) + \lambda \left[\sigma^2(x){+}
  {\vec{\pi}}^2(x) {-} v'^2(x) \right] \vec{\pi}(x)
  &=&\,- g \vec\rho_P(x),\nonumber\\
  \label{EoM}
\end{eqnarray}
where $v'^2(x) \equiv v^2-T^2(x)/2$.
Here $\rho_S \equiv \langle \bar{q} q\rangle $ and $\vec\rho_P
\equiv i\langle
\bar{q}\gamma_5
\vec{\tau} q\rangle$ are the scalar and pseudoscalar quark densities,
which
should be determined self-consistently from the
motion of $q$ and $\bar{q}$ in the background meson fields.
The scalar and pseudoscalar densities can be
represented as \cite{csemis95}
\begin{equation}
   \rho_S(x) = g a(x)\sigma(x), ~~~\vec\rho_P(x) = g a(x)\vec{\pi}(x),
   \label{den}
\end{equation}
where $a(x)$ is expressed in terms of the
momentum distribution of quarks and antiquarks $f(x, p)$:
$$
  a(x) = \nu_q \int \frac{d^4p}{(2\pi\hbar)^3}\
  2\delta(p^\mu p_\mu - m^2(x)) f(x, p)
$$
\begin{equation}
   \longrightarrow\frac{\nu_q}{(2\pi\hbar)^3} \int \frac{d^3p}{E(x, {\bf
   p})}
   [n_q(x, {\bf p}) + n_{\bar{q}}(x, {\bf p})] \ .
   \label{e3}
\end{equation}
Here $\nu_q$ is the degeneracy factor of quarks 
%%%%%%%%%%%%%%%%%%%%%%
% initial submission %
%which equals to 12 in our case. 
%%%%%%%%%%%%%%%%%%%%%%
% 1st revision       %
which equals to 12 in our case
(flavour$\;\times\;$spin$\;\times\;$colour).
%%%%%%%%%%%%%%%%%%%%%%

%%%%%%%%%%%%%%%%%%%%%%
% initial submission %
%Let us consider a one-dimensional Bjorken scaling, 
%i.e., that $\sigma$, $\vec\pi$ and $a$ are functions of the proper
%time only. 
%Let us assume 
%%%%%%%%%%%%%%%%%%%%%%
% 1st revision       %
Let us consider a one-dimensional Bjorken scaling and assume 
%%%%%%%%%%%%%%%%%%%%%%
that
the quark distribution is an ideal Fermi
distribution at the freeze-out time $\tau_0$ 
(following
\cite{csocse94,csemis95} we can take $\tau_0 \sim 7{\rm fm}/c$)
%%%%%%%%%%%%%%%%%%%%%%
% initial submission %
%and 
%%%%%%%%%%%%%%%%%%%%%%
% 1st revision       %
while
%%%%%%%%%%%%%%%%%%%%%%
the subsequent distribution for time $\tau > \tau_0$ is given
by the solution of the collisionless Vlasov equation \cite{csemis95}.
%%%%%%%%%%%%%%%%%%%%%%
% initial submission %
%Then the source term, $a(x)$, in the  EoM
%%%%%%%%%%%%%%%%%%%%%%
% 1st revision       %
Then the source term, $a(x)$,
%%%%%%%%%%%%%%%%%%%%%%
can be cast into \cite{fenmol97}: 
\begin{eqnarray}
  a(x)&\equiv& a(\tau)=\frac{\nu_q\alpha}{\pi^2\sqrt{1-\alpha^2}}\;T_0^2 
   \nonumber\\
  &\times&
  \int\limits_0^\infty dq\;q\; 
  \frac{
    {\rm arcsin}\sqrt{
    \left( 1-\alpha^2 \right)/\left( \frac{m^2(\tau)}{T_0^2 q^2}+1
    \right) } 
    }
    {
    \exp\left( 
   \sqrt{q^2+m^2(\tau_0)/T_0^2}
              -\frac{\mu_0}{T_0} \right)+1} 
   \ , \nonumber
\end{eqnarray}
where $\alpha\equiv\tau_0/\tau$, $T_0$ and $\mu_0$ are the freeze-out
temperature and chemical potential.
For the $\mu_0=0$,  $m_q=0$
case this expression can be integrated analytically \cite{csemis95},  
while in the general $m_q\neq 0$, $\mu_0\neq 0$ case it 
can be evaluated numerically using an
eight-point
Gauss--Laguerre integral formula \cite{fenmol97}.  This way we can follow
the
solution of the EoM numerically from an arbitrary
initial condition at $\tau_0$ during the whole symmetry breaking process.
For simplicity, we shall assume that quarks and antiquarks are in
chemical
equilibrium, i.e., we take $\mu_0=0$.

To solve the EoM, we need to know the temperature field, $T(x)$. 
However, following \cite{rajwil93,rajwil93_2} we may assume that
%\cite{biaczy95,huawan94,csemis95,csemis96,birmol97,fenmol97}.
the cooling is very fast
and then we may take $T(x)\equiv 0$. This is the so-called {\em quench}
scenario. Certainly, the quench is an idealization since in a real collision
thermal fluctuations will not cease
instantaneously.
Now 
the EoM has
no reference to any temperature but 
the initial condition, which we choose to be at the freeze-out of the
quarks, must
represent the freeze-out temperature, $T_0\approx 130-140{\rm MeV}$. 
We emphasize that our scenario differs from the original quench because, 
even though the thermal fluctuations of the fields are neglected, we
still have an interaction with the expanding quark/antiquark background. 
One could choose other possible non-equilibrium scenarios,
for example, {\em annealing} \cite{gavgoc94,asahua95}.

Before the quench, the system can typically be found somewhere around the
minimum of
the potential energy density.
However, 
%%%%%%%%%%%%%%%%%%%%%%
% initial submission %
%when we perform the quench,
%%%%%%%%%%%%%%%%%%%%%%
% 1st revision       %
when the quench occurs,
%%%%%%%%%%%%%%%%%%%%%%
the position of this minimum suddenly
changes to the
physical vacuum, $(\sigma,\vec\pi)=(f_\pi,\vec 0)$.
Thus, the system finds itself on the slope of the "valley"
of
the potential,  
rolls down towards its new minimum, and oscillates around it
\cite{gavmul94,huawan94,S96,fenmol97}.
The
oscillations are damped
%%%%%%%%%%%%%%%%%%%%%%
% initial submission %
%because of the presence of a first order
%derivative, $\partial/\partial_\tau$,  
%%%%%%%%%%%%%%%%%%%%%%
% 1st revision       %
because, 
due to the expansion, 
there is a first order derivative
%%%%%%%%%%%%%%%%%%%%%%
in the EoM (\ref{EoM}), therefore
the
fields
finally relax to
the physical vacuum, $(f_\pi,\vec 0)$.

Studies show \cite{S96,fenmol97}
that the main effect of the quarks is an additional damping;
the oscillations of the chiral fields are reduced by about $40-50$\%.
This is due to the
fact that during the
transition a portion of the energy of the system is used up for
the increasing quark mass. 
Also, the phase transition happens about 0.5fm/$c$ slower; it takes about
2.5fm/$c$.
The presence of quarks has a strong influence on the onset of
instabilities and domain formation in the initially homogeneous system.
This we shall address in a separate paper.
% ###D.: Will we?           ^^^^^^^^^^^^
% - Oct 21, 1997

\section{The question of the initial condition}
\label{Ch:Init}

The physical choice of initial condition or ensembles of initial states
was not studied in detail in earlier works, although this is a vital problem if
we want to predict the observability of DCC fluctuations.
An early study \cite{huawan94} followed by refs. \cite{S96,fenmol97}
assumed the initial condition:
\begin{eqnarray}
  &&\sigma_0=0, \ \vec\pi_0=\vec 0,\  \nonumber\\
  &&\dot{\sigma_0}=1{\rm
MeV}\,c/{\rm fm}, \   
  \dot{\vec\pi_0}=(5,0,0){\rm MeV}\,{c}/{\rm fm}, 
  \label{usinit}
\end{eqnarray}
with $\tau_0=1{\rm fm}/c$.
This choice of the chiral field configuration
corresponds to the chirally symmetric phase representing very hot
quark-gluon plasma, i.e., when $H=0$ and 
%%%%%%%%%%%%%%%%%%%%%%
% initial submission %
%$T>T_c\equiv \sqrt{v}$ 
%%%%%%%%%%%%%%%%%%%%%%
% 1st revision       %
$T>T_c = \sqrt{2} v$ 
%%%%%%%%%%%%%%%%%%%%%%
($\approx124$MeV). 
The initial value of the fields was assumed to be the value of the
condensate, which is zero in this case, i.e., $\langle\sigma\rangle=0$ and
$\langle\vec\pi\rangle=0$.

\begin{figure}[htb]
\vspace*{-0.5cm}
\center
\leavevmode
\epsfysize=6cm
\epsfbox{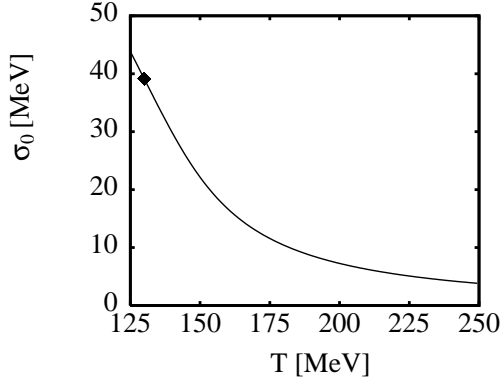}
       \vspace*{-0.5cm}
%%%%%%%%%%%%%%%%%%%%%%
% initial submission %
%\caption[Sigma condensate as a function of temperature in the case of
%explicitly broken chiral symmetry]
%{\footnotesize
%The $\sigma$ condensate as a function of the temperature,
%%%%%%%%%%%%%%%%%%%%%%
% 1st revision       %
\caption []
{\footnotesize
The minimum of the potential energy density 
as a function of the temperature
%%%%%%%%%%%%%%%%%%%%%%
with
our standard model parameters. 
Clearly, the condensate is non-zero even at
high temperatures.
The 'diamond' indicates its value 
at the freeze-out temperature $T_0=130$MeV.}
       \label{fig:T-cond}
\end{figure}

However, to 
get the correct value of the 
pion mass in the  physical vacuum, we do need the explicit symmetry
breaking term $H\sigma$. Hence, in the minimum of the potential the
condensate
is never zero, though it gets close to zero for temperatures above
$250-300$MeV
(see Fig. \ref{fig:T-cond}). However, at the freeze-out temperature,
$T_0=130$MeV, we have a large condensate of 40MeV.

Another important factor is that our
system is finite, therefore thermal fluctuations are of real importance.
Due to the fluctuations, the system is at the minimum of the potential
energy density
only {\em on the
average}, and the probability for finding it somewhere around this minimum
can be considerable, especially for very small systems and/or high
temperatures.
Therefore, the time evolution starting from the minimum of the potential
energy density is not fully representative for the majority of the
realistic initial states.   

Furthermore, the system has a considerable kinetic energy at the
temperatures
under
consideration. This should be reflected by the choice of the initial value of 
the field derivatives. Our
analysis will show that the initial condition
(\ref{usinit}) has too little kinetic energy for the temperatures we are
considering, even if we go down to as low as $T=130$MeV. Thus,
the picture of
a system on the "top" of the potential started with a
little "kick" cannot be realistic. 

These considerations lead us to propose an initial condition, which is
closer to the physical picture we have at the time of the freeze-out.
Before the freeze-out the system can be approximated to be 
in thermal equilibrium at a given
temperature, $T$, and it has a given volume, $V$. 
If so, then
the initial condition
should be chosen according to a canonical thermal ensemble, i.e., the
probability $p$ of a configuration $(\sigma,\vec\pi, \dot\sigma,
\dot{\vec\pi})$ is proportional to the
Boltzmann-factor, $P$:
\begin{eqnarray}
  p(\sigma,\vec\pi, \dot{\sigma},\dot{\vec\pi})&\propto&
  \exp\left[-\frac{V}{T}{\cal H}(\sigma,\vec\pi, \dot{\sigma},
  \dot{\vec\pi})\right] \nonumber\\
  &\equiv& P(\sigma,\vec\pi, \dot{\sigma},\dot{\vec\pi})
  \label{candist}
\end{eqnarray}
where ${\cal H}$ is the Hamiltonian density, the dot stands for
$\partial/\partial\tau$, and
we assumed that the
system
is homogeneous and we have taken the Bjorken scaling granted.
The
Hamiltonian density consists of three terms, ${\cal H}_{kin}$, ${\cal
H}_{pot}$
and ${\cal H}_q$, which are respectively the kinetic and potential energy
density of the 
chiral fields and the energy density of the quarks (with $\mu=0$):
\begin{eqnarray}
%%%%%%%%%%%%%%%%%%%%%%
% initial submission %
%   {\cal H}_q&=&\nu_q\int \frac{d^3p}{E}\, E\, 
%   \left[f_q(x,\vec p)+f_{\bar q}(x,\vec p)\right] \nonumber\\
%   &=&
%   \frac{2\nu_q}{\left(2\pi\right)^3}
%   \int \frac{d^3p}{\exp\left[\sqrt{\vec p^2+m^2}/T\,\right]+1}
%          \nonumber \\
%%%%%%%%%%%%%%%%%%%%%%
% 1st revision       %
   {\cal H}_q&=&\nu_q\int \frac{d^3p}{E}\, E^2\, 
   \left[f_q(x,\vec p)+f_{\bar q}(x,\vec p)\right] \nonumber\\
   &=&
   \frac{2\nu_q}{\left(2\pi\right)^3}
   \int d^3p\ \frac{\sqrt{\vec p^2 + m^2}}
        {\exp\left[\sqrt{\vec p^2 + m^2}/T\,\right]+1}
   \nonumber \\
%%%%%%%%%%%%%%%%%%%%%%
   {\cal H}_{kin}&=&\frac{1}{2}\dot\sigma^2
               +\frac{1}{2}\dot{\vec\pi}^2 
                         \nonumber \\
   {\cal H}_{pot}&=&U(\sigma,\vec\pi)=\frac{\lambda}{4}
          \left(\sigma^2+\vec\pi^2-v'^2\right)^2-H\sigma
          + U_0
         \ ,
   \label{endens}
\end{eqnarray}
where the effective quark (antiquark) mass is given 
as $m^2=g^2(\sigma^2+\vec\pi^2)$.

Similar approaches can be found in the literature\ \ 
where Gaussian distributions were
considered \ \  
\cite{rajwil93_2,gavgoc94,asahua95,huawan94}.
However, as
the potential energy density is not
qua\-drat\-ic and we also have the quark term, the distribution of the
chiral fields is clearly
non-Gaussian. 
%%%%%%%%%%%%%%%%%%%%%%%%%%%
% 1st revision - inserted %
We shall give a comparison 
to the initial condition of refs. \cite{rajwil93_2,gavgoc94}:
\begin{equation}
  \langle \phi \rangle = \langle\dot\phi \rangle = 0, \; 
  \langle \phi^2 \rangle = v^2/4, \; \langle \dot\phi^2 \rangle =
  v^2/1{\rm fm}^2.
\label{gaussinit}
\end{equation}
%%%%%%%%%%%%%%%%%%%%%%%%%%%

A forerunner of the present analysis is ref. \cite{birmol97} where we
considered
initial conditions with large chiral "angular momenta", $I$, i.e., states
where the chiral fields circle around the
minimum of the
potential energy density having a trajectory which is far from the
minimum. 
As we have
shown
there, these states as well lead to DCC formation during the fast linear
expansion. 
However, the magnitude of $I$ remained an
open question. Now we shall extend the analysis by including the
initial distribution of finite chiral 'angular momentum' states.

\begin{figure}[htbp]
\center
\leavevmode
\epsfysize=6cm
\epsfbox{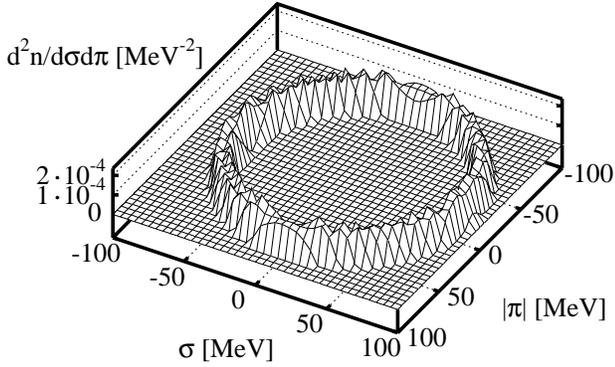}
       \vspace*{0cm}
%%%%%%%%%%%%%%%%%%%%%%%
% initial submission %
%\caption[Initial distribution of the chiral fields $\sigma$ and $\pi_1$, 
%for the microcanonical distribution in Fig. 7 in ref.
%\protect\cite{birmol97}] 
%%%%%%%%%%%%%%%%%%%%%%%
% 1st revision        %
\caption[]
%%%%%%%%%%%%%%%%%%%%%%%
{\footnotesize 
The initial distribution 
%%%%%%%%%%%%%%%%%%%%%%
% initial submission %
%of the chiral fields $\sigma$ and $\pi_1$, 
%%%%%%%%%%%%%%%%%%%%%%
% 1st revision       %
of the chiral field $\sigma$ and the magnitude of the isovector $\vec\pi$,
%%%%%%%%%%%%%%%%%%%%%%
for
the microcanonical distribution in Fig. 7 in ref. \protect\cite{birmol97}.
Unlike the peaked canonical distributions in Fig.
\protect\ref{fig:sp_2-10-50}, the
states are concentrated 
%%%%%%%%%%%%%%%%%%%%%%
% initial submission %
%around circles. 
%%%%%%%%%%%%%%%%%%%%%%
% 1st revision       %
in a circular band.  
The initial ensemble contained 600,000 states and, 
to aid the eye, 
the distribution was complemented by a reflection 
on the $|\vec\pi|=0$ plane. 
%%%%%%%%%%%%%%%%%%%%%%
}
       \label{fig:birmol97:fig7}
\end{figure}

There are several methods to generate the ensemble of the initial
conditions.
In ref. \cite{birmol97} we needed a {\em microcanonical} ensemble 
around a specified value of the "angular momentum",
$I$. This ensemble was obtained by sampling the time evolution of the
fields evenly in the case with explicit symmetry breaking but
without quarks and
expansion (Fig. \ref{fig:birmol97:fig7}).  In this case $I$ is not a
constant of motion but oscillates in a certain interval. This method is
based on the assumption of the
ergodicity of the
system, i.e., that during the time evolution the system gets arbitrarily
close to any point in the available phase space. 
%%%%%%%%%%%%%%%%%%%%%%
% initial submission %
%As in this work we have a {\em canonical} distribution to generate,
%we had to choose another method, namely the Monte-Carlo method.
%Consequently, we shall use a different formalism from that in ref.
%\cite{birmol97}.
%%%%%%%%%%%%%%%%%%%%%%
% 1st revision       %
It is also possible to generate a {\em canonical} ensemble 
based on ergodicity. 
In ref. \cite{birgre97} a Langevin type equation of motion 
was used to generate such initial states, 
where a connection to an external heatbath of a given temperature 
was assumed. 
Then, based on the ergodicity, 
the states at different times correspond to 
a canonical ensemble of the system. 

In this work we generate our initial canonical ensemble 
using a different method, 
namely the Monte--Carlo method. 
Consequently, 
we shall use a different formalism from that 
in ref. \cite{birmol97} or ref. \cite{birgre97}.
%%%%%%%%%%%%%%%%%%%%%%

\subsection{The generation of the canonical distribution}

To generate the canonical distribution (\ref{candist}), we used a
simple generalization of the so-called rejection method \cite{NumRec}
for a multidimensional probability distribution on a boundless manifold.
A convenience of our method is that it does not require proper
normalization of the
probability distribution, i.e., the evaluation of the normalization
factor.

To apply the method we need to find a function, 
$F(\sigma,\vec\pi,\dot{\sigma},\dot{\vec\pi})$, 
which is everywhere above 
$P(\sigma,\vec\pi,\dot{\sigma},\dot{\vec\pi})$ 
and is proportional to a
distribution, $f(\sigma,\vec\pi, \dot{\sigma},\dot{\vec\pi})$, which we
are
able to generate.
It is simple to prove that
if one generates a (8-dimensional) random deviate, $(\sigma,\vec\pi,
\dot{\sigma},\dot{\vec\pi})$, 
of $f$ and a uniform
deviate
$y$ between 0 and $F(\sigma,\vec\pi, \dot{\sigma},\dot{\vec\pi})$, 
then the $(\sigma,\vec\pi, \dot{\sigma},\dot{\vec\pi},y)$ points 
will be uniform in the (9-dimensional) volume under the graph of $F$.
If now 
one keeps the point $(\sigma,\vec\pi, \dot{\sigma},\dot{\vec\pi},y)$
whenever it is under $P$ and rejects it if
not, the accepted $(\sigma,\vec\pi, \dot{\sigma},\dot{\vec\pi})$ points
will have the desired
$p(\sigma,\vec\pi, \dot{\sigma},\dot{\vec\pi})$
distribution.  
The fraction of accepted points, i.e., the efficiency of the method is 
$$
  A=\frac{\int\cdots\int F(\sigma,\vec\pi, \dot{\sigma},\dot{\vec\pi}) 
 \, d\sigma\, d^3\pi\, d\dot\sigma\, d^3\dot\pi}
  {\int\cdots\int P(\sigma,\vec\pi, \dot{\sigma},\dot{\vec\pi}) 
 \, d\sigma\, d^3\pi\, d\dot\sigma\, d^3\dot\pi} \ .
$$

To make it easy to
generate the $f$ distribution corresponding to $F$, we
shall choose
$F$ to be proportional to a product of Gaussian deviates:
\begin{eqnarray}
   F(\sigma,\vec\pi, \dot{\sigma},\dot{\vec\pi})&=&C
   e^{-C_1\left(\sigma-\sigma_m\right)^2}  
   e^{-C_1\vec\pi^2}  
   e^{-C_2{\dot\sigma}^2}  
   e^{-C_2{\dot{\vec\pi}}^2}\nonumber\\
   &\equiv& F_1(\sigma,\vec\pi) \times F_2(\dot\sigma,\dot{\vec\pi}).
\end{eqnarray}
Now $f$ will simply factorize into one-dimensional Gaussian distributions
of the chiral fields and their derivatives. These Gaussian deviates can
be generated using the routine {\em gasdev} 
described in ref. \cite{NumRec}.
The random number generator we
used is the
routine 
{\em ran2} in ref. \cite{NumRec}.

%%%%%%%%%%%%%%%%%%%%%%
% initial submission %
%\subsubsection{The choice of the distribution function}
%
%We have to choose the coefficients, $C_1$ and $C_2$, in $F$ such that 
%%%%%%%%%%%%%%%%%%%%%%
% 1st revision       %
\subsubsection{The choice of the comparison function}

We have to choose the parameters, $C$, $C_1$, $C_2$ and $\sigma_m$, 
in $F$ such that 
%%%%%%%%%%%%%%%%%%%%%%
\begin{equation}
  F(\sigma,\vec\pi,\dot\sigma,\dot{\vec\pi}) \leq
  P(\sigma,\vec\pi,\dot\sigma,\dot{\vec\pi})
  \label{ineqcomp}
\end{equation} 
for all
values of the chiral fields and their derivatives.

The coefficient $C_2$ is determined by the kinetic energy density
term because the field derivatives appear only there. Since it is a
quadratic function of the derivatives, 
it is straightforward to choose $C_2=V/2T$, i.e.,
\begin{equation}
  F_2(\dot\sigma,\dot{\vec\pi})=
  \exp\left[-\frac{V}{2T}\left(\dot\sigma^2+\dot{\vec\pi}^2\right)\right].
  \label{F2}
\end{equation}

The fields are in the other two terms of the Hamiltonian density, which
are not quadratic, therefore we 
%%%%%%%%%%%%%%%%%%%%%%
% initial submission %
%have to use some kind of 
%an approximation 
%to choose $C_1$.
%%%%%%%%%%%%%%%%%%%%%%
% 1st revision       %
need approximations.
%%%%%%%%%%%%%%%%%%%%%%
With the help of Eq. (\ref{F2}), inequality (\ref{ineqcomp}) can be
written as 
$$
  \frac{V}{T}\left({\cal H}_{pot}+{\cal H}_q\right)\ge
  C_1\left(\sigma-\sigma_m\right)^2+C_1 \vec\pi^2+const.,
$$ 
thus we have to approximate the potential and quark energy density terms
with a qua\-drat\-ic expression from {\em
below}.

%%%%%%%%%%%%%%%%%%%%%%
% initial submission %
%The quark energy density contains the fields in a rather intractable
%integral. 
%However, 
%the quark energy density is in the range between 0 and 0.627
%MeV/fm$^3$ 
%for our standard parameters,
%which is much smaller than the changes in the potential and kinetic
%energy
%density. Therefore, it is 
%a very good approximation of the quark energy
%density from below to take the constant value  as zero.
%%%%%%%%%%%%%%%%%%%%%%
% 1st revision       %
The quark energy density contains the fields 
in a rather intractable integral. 
It is a monotone decreasing function of the quark mass and it is in the
range between 0 and 257MeV/fm$^3$.  
It turns out to be
sufficient to take a simple, crude lower bound  
of the form $L - M m^2$, where $M > 0$.
%%%%%%%%%%%%%%%%%%%%%%

It is enough to deal with the quartic term in the potential
energy
density (\ref{endens}), since the remaining $H\sigma$ term can
easily be combined
with 
any quadratic expression of $\sigma$ 
into a final $C_1\left(\sigma-\sigma_m\right)^2$ 
quadratic expression. 
The
quartic term has the general form
%%%%%%%%%%%%%%%%%%%%%%
% initial submission %
%$$
%  C'\left(a^2\pm 1\right)^2,  
%$$
%%%%%%%%%%%%%%%%%%%%%%
% 1st revision       %
$C'\left(a^2\pm 1\right)^2$,
%%%%%%%%%%%%%%%%%%%%%%  
where $a^2\equiv\left(\sigma^2+\vec\pi^2\right)/|v'^2|$, 
 the upper sign $(+)$ corresponds to $T>T_c$ while the
lower one $(-)$
to $T<T_c$, and $C'$ does not contain the fields. We approximate this
expression
by a quadratic one 
using that
%%%%%%%%%%%%%%%%%%%%%%
% initial submission %
%$$
%  \left(a^2\pm 1\right)^2\ge la^2-k, \qquad \qquad (l\ge 0)
%$$
%%%%%%%%%%%%%%%%%%%%%%
% 1st revision       %
$\left(a^2\pm 1\right)^2\ge la^2-k$,
%%%%%%%%%%%%%%%%%%%%%%
if 
\begin{equation}
   0\le l\le 2\sqrt{k+1}\pm 2.
   \label{ineqlk}
\end{equation}

Putting everything together, the
first part of the comparison function is 
%%%%%%%%%%%%%%%%%%%%%%
% initial submission %
%\begin{eqnarray}
% F_1(\sigma,\vec\pi)&=&
%\exp\left\{-\frac{V}{T}\frac{\lambda^2 l |v'^2|}{4}
%     \left[\left(\sigma-\frac{2H}{\lambda^2 l |v'^2|}\right)^2
%     +\vec\pi^2\right]\right\}\nonumber\\
%     &\times& \exp\left(
%     \frac{V}{T}\frac{H^2}{\lambda^2 l |v'^2|}
%     +\frac{V}{T}\frac{\lambda^2 v'^4 k}{4} \right).
%\end{eqnarray}
%where, following from the definition of $v'^2$, 
%$$
%  |v'^2|\equiv\left\{\matrix{v^2-T^2/2 \qquad {\rm below}\ T_c \cr
%                           T^2/2-v^2 \qquad {\rm above}\ T_c} \right. \ .
%$$  
%%%%%%%%%%%%%%%%%%%%%%
% 1st revision       %
\begin{eqnarray}
 F_1(\sigma,\vec\pi)&=&
\exp\left\{-\frac{V}{T}\frac{\lambda^2 l' |v'^2|}{4}
     \left[\left(\sigma-\frac{2H}{\lambda^2 l' |v'^2|}\right)^2
     +\vec\pi^2\right]\right\}\nonumber\\
     &\times& \exp\left[\frac{V}{T}
      \left(\frac{H^2}{\lambda^2 l' |v'^2|}
     +\frac{\lambda^2 v'^4 k}{4} - L \right) \right] \ ,
\end{eqnarray}
with $l' \equiv l - 4Mg^2/\lambda^2|v'^2|$.
%%%%%%%%%%%%%%%%%%%%%%

%%%%%%%%%%%%%%%%%%%%%%
% initial submission %
%It is crucial to choose the parameters $l$ and $k$ such that we have a
%reasonable efficiency
%(otherwise it may even be as small as $10^{-6}$!).
%Clearly, the best choice is the one, which minimizes 
%the area under $F_1$ since that gives the highest efficiency. 
%It is easy to see that for a fixed $l$ we need $k$ to be minimal, thus
%we have to take the upper limit in Ineq. (\ref{ineqlk}), i.e.,
%$l=2\sqrt{k+1}\pm 2$. Now we have only one variable, $k$, left which can
%be
%fixed by evaluating the area under $F_1$ and looking for the extremal
%points. This leads to a third-order algebraic equation for $k$, 
%which we solved numerically. Using this
%optimization, 
%we obtained an efficiency of about 20\%.
%%%%%%%%%%%%%%%%%%%%%%
% 1st revision       %
It is crucial to choose the parameters $l$, $k$, $L$ and $M$ 
such that we have a reasonable efficiency.
Clearly, 
to get the highest efficiency we have to minimize the area under $F_1$.
It is easy to see that for a fixed $l$ we need $k$ to be minimal, 
thus we have to take the upper limit in Ineq. (\ref{ineqlk}),
$l=2\sqrt{k+1}\pm 2$. 
Since the quark energy density is a convex function of $m^2$, 
for a given $M$ the optimal, i.e., the largest $L$ is given 
by the tangent line that has a slope $(-M)$.
Now we have only $k$ and $M$ left. 
For a given $M$, 
the minimum condition for the area under $F_1$ leads 
to a third-order algebraic equation for $k$ that we solved numerically.
$M$ we set by trial and error.
Having the above optimization we obtained an efficiency of 30---60\%.
%%%%%%%%%%%%%%%%%%%%%%

The quark energy can be expressed as \cite{Csernai94}
%%%%%%%%%%%%%%%%%%%%%%
% initial submission %
%\begin{equation}
%   {\cal H}_q=\frac{\nu_q}{\pi^2} T m^2 \sum_{k=1}^\infty (-1)^{k-1}
%     \;\frac{1}{k} K_2\left(\frac{km}{T}\right),
%\end{equation}
%where $K_2$ stands for the modified Bessel function of the
%second kind of second order. 
%%%%%%%%%%%%%%%%%%%%%%
% 1st revision       %
$$
   {\cal H}_q=\frac{\nu_q}{2 \pi^2} T m^2 \sum_{k=1}^\infty (-1)^{k-1} 
   \left[ \frac{m}{k} K_3\!\!\left(\frac{km}{T}\right) 
     - \frac{T}{k^2} K_2\!\!\left(\frac{km}{T}\right) \right],
$$
where $K_n$ stands for the modified Bessel function of the
second kind of $n$-th order.
%%%%%%%%%%%%%%%%%%%%%%
Certainly, we cannot take into account all
terms here but that is unnecessary as the sum is convergent. We summed up
the terms until the relative
change of the sum became less than $10^{-6}$ (this corresponds to $4-16$
terms). 

The above formula cannot be used directly for $m=0$ and for $T=0$.
For $m=0$ we can obtain the expression \cite{Csernai94}
%%%%%%%%%%%%%%%%%%%%%%
% initial submission %
%$$
%   {\cal H}_q=\frac{3\nu_q}{2\pi^2}T^3\zeta(3) \ ,
%$$
%where $\zeta$ is the Riemann zeta function ($\zeta(3)\approx 1.20206$).
%%%%%%%%%%%%%%%%%%%%%%
% 1st revision       %
$
   {\cal H}_q=7\pi^2\nu_q T^4/120\ .
$
%%%%%%%%%%%%%%%%%%%%%%
For $T=0$ the integrand and thus the quark energy density is zero (see
ref.
\cite{Csernai94}, Section 7.3).

\subsection{An advantage of the symmetries of the Hamiltonian}
\label{Ssc:advsym}

%%%%%%%%%%%%%%%%%%%%%%
% initial submission %
%Usually it is the easier task to generate the ensemble of the proper
%initial condition. The more difficult is to run the calculations since we
%may
%need several thousand initial states in the ensemble. 
%%%%%%%%%%%%%%%%%%%%%%
% 1st revision       %
Though it does not take long 
to generate the ensemble of the proper initial conditions, 
it is very time-consuming
to compute the time evolution from the several thousand initial states. 
%%%%%%%%%%%%%%%%%%%%%%
However, 
it is possible to exploit the symmetries of the Hamiltonian to reduce the
computation.

%%%%%%%%%%%%%%%%%%%%%%
% initial submission %
%Let us take a look at the EoM (\ref{EoM}). These equations
%have
%the following symmetry: if we introduce a new vector $\vec\pi'$ by
%rotating the $\vec\pi$ fields with an arbitrary rotation matrix ${\bf
%O}$,
%the
%equations will have the same form but for $\vec\pi'={\bf O} \vec\pi$.
%%%%%%%%%%%%%%%%%%%%%%
% 1st revision       %
It is easy to verify the following symmetry: 
if a trajectory 
$\left(\sigma\left(\tau\right), \vec\pi\left(\tau\right)\right)$
satisfies the EoM (\ref{EoM}), 
so does the trajectory 
$\left(\sigma\left(\tau\right), {\bf O}\vec\pi\left(\tau\right)\right)$, 
where ${\bf O}$ is an arbitrary, 
$3\times 3$ rotation matrix in isospin space.
%%%%%%%%%%%%%%%%%%%%%%
Obviously, the rotation implies that the initial condition as well should
be rotated.
Now, if we have data for a given initial condition, we  at once know the
results for all other initial conditions obtained from this given one by
rotations; we simply have to rotate the pion fields using the same
rotations. 

%%%%%%%%%%%%%%%%%%%%%%
% initial submission %
%Therefore, in this sense all initial conditions, which are
%rotationally equivalent are the same.
%%%%%%%%%%%%%%%%%%%%%% 
% 1st revision     %
In this sense all rotationally equivalent initial conditions are the same.
%%%%%%%%%%%%%%%%%%%%%%
We may use this to reduce the number
of parameters in the initial condition. 
Originally these are 8 (four fields +
four derivatives). However, taking into account the rotations we may fix
two pion fields and the sign of the third one, e.g., we can take
$\pi_1=\pi_2=0$ and $\pi_3\ge 0$, and we can also fix one derivative,
e.g.,
 $\dot {\pi_2}=0$. Thus, we have only 5 parameters left.

%%%%%%%%%%%%%%%%%%%%%%
% initial submission %
%The above parametrisation can be reexpressed in a form 
%which is more convenient to generate the
%canonical distribution. 
%The three new parameters are  
%%%%%%%%%%%%%%%%%%%%%%
% 1st revision       %
Since our canonical distribution contains the pion fields only 
through $\vec\pi^2$ and $\dot{\vec\pi}^2$, 
it is convenient to reexpress the above parametrisation using
%%%%%%%%%%%%%%%%%%%%%%
the length of the isovector,
$|\vec\pi|$, the length of the derivative, $|\dot{\vec\pi}|$, and the
angle, $\theta$, between $\vec\pi$ and $\dot{\vec\pi}$. Their relation
to the previously used parameters is trivial: $\pi_3=|\vec\pi|$,
$(\dot{\pi_3}, \dot{\pi_1})=|\dot{\vec\pi}| (\cos\theta,\sin\theta)$.
%%%%%%%%%%%%%%%%%%%%%%
% initial submission %
%The convenience
%comes from the fact that the canonical distribution contains the pion
%fields
%only through $|\vec\pi|$ and $|\dot{\vec\pi}|$, thus 
%%%%%%%%%%%%%%%%%%%%%%
% 1st revision       %
Now
%%%%%%%%%%%%%%%%%%%%%%
we may realize the
distribution in spherical coordinates:
\begin{eqnarray}
  &\exp&\left[-\frac{V}{T}{\cal H}(\sigma, \vec\pi, \dot\sigma,
  \dot{\vec\sigma})\right]\, d\sigma\, d\vec\pi\, d\dot\sigma\,
  d\dot{\vec\pi}= \nonumber\\
  &\exp&\left[-\frac{V}{T}{\cal H}(\sigma, |\vec\pi|, \dot\sigma,
  |\dot{\vec\pi}|)\right]\, \vec\pi^2 \dot{\vec\pi}^2
   d\sigma\, d|\vec\pi|\, d\dot\sigma\,   d|\dot{\vec\pi}|\;\nonumber\\
   &&\times\sin\Theta\, d\Theta\, d\Phi\; \sin\vartheta\, d\vartheta\,
d\varphi \ ,
  \label{candistpolar}
\end{eqnarray}
where the angles $\vartheta$ and $\varphi$ correspond\footnote{Here, we 
mean the usual parametrisation
%%%%%%%%%%%%%%%%%%%%%%
% initial submission %
%\begin{eqnarray}
%   \pi_1&=&|\vec\pi|\,\sin\vartheta\, \cos\varphi\ , \nonumber\\
%   \pi_2&=&|\vec\pi|\,\sin\vartheta\, \sin\varphi\ , \nonumber\\
%   \pi_3&=&|\vec\pi|\,\cos\vartheta \nonumber \ ,
%\end{eqnarray}
%%%%%%%%%%%%%%%%%%%%%%
% 1st revision       %
$\pi_1\equiv|\vec\pi|\,\sin\vartheta\, \cos\varphi$,
$\pi_2\equiv|\vec\pi|\,\sin\vartheta\, \sin\varphi$,
and $\pi_3\equiv|\vec\pi|\,\cos\vartheta \nonumber$,
%%%%%%%%%%%%%%%%%%%%%%
with $\varphi\,\epsilon\,[0,2\pi)$ and $\vartheta\,\epsilon\,[0,\pi)$;
and a similar one for the vector $\dot{\vec\pi}$.}
to $\vec\pi$, while the other two angles, $\Theta$ and $\Phi$,
correspond to  $\dot{\vec\pi}$. The orientation of the two spherical
frames can be chosen arbitrarily. We take the 
choice when $\vec\pi$ is along the $z$-axis in the
frame of $\dot{\vec\pi}$ because then $\Theta$ is the angle
between these two vectors. Exploiting the rotational invariance,  
the angles $\varphi,\ \vartheta$ and $\Phi$ can be fixed, e.g., to
$\varphi=\vartheta=\Phi=0$, and
the 5 parameters needed for the initial condition are $\sigma,\
|\vec\pi|,\ 
\dot\sigma,\ |\dot{\vec\pi}|$ and $\Theta$.
%%%%%%%%%%%%%%%%%%%%%%
% initial submission %
%The distribution of the three angles,
%$\varphi$, $\vartheta$ and $\Phi$, will then be needed when we want 
%%%%%%%%%%%%%%%%%%%%%%
% 1st revision       %
We will only need the distribution of the three angles,
$\varphi$, $\vartheta$ and $\Phi$, 
%%%%%%%%%%%%%%%%%%%%%%
to reconstruct the complete, 8-variable 
distributions of any physical quantity from the results for the ensemble
of the 5-variable initial conditions.

The distribution (\ref{candistpolar}) factorizes 
into three terms: a term depending on the fields and field derivatives,
$\sigma$, $\dot\sigma$, $|\vec\pi|$ and $|\dot{\vec\pi}|$,
a term depending on
$\Theta$, and a third term containing the other three angles, $\Phi$,
$\vartheta$ and $\varphi$. We may
generate the first term using the rejection method introduced in the
previous subsection. The second term can be obtained by transformation
from a
uniform deviate (see ref. \cite{NumRec}). This way, we can have a
representative ensemble of the 5 parameters needed for the initial
conditions. We can easily obtain the ensemble for the remaining
three angles, for example, using the rejection method again. By forming
pairs from the elements of the two ensembles,
(let us define this operation as the product of the two ensembles) we
have a representative ensemble of the original canonical distribution.
However, we need to run the calculations only for the first ensemble in
the product, while the second one we {\em apply} on the results as
rotations. This
enables us to choose the second ensemble as large as we wish. Furthermore,  
there are parameters, for example the neutral pion ratio, that we can 
calculate {\em analytically} by
taking
into account the distribution function of the second ensemble!

Let us show this in more detail. The joint distribution of the angles
$\varphi$ and 
$\vartheta$ is independent of the other
variables in (\ref{candistpolar}), moreover, it is the uniform
distribution on the three-sphere, 
given by \ $\sin\vartheta\, d\vartheta\,
d\varphi$. 
To reconstruct the distribution of the $\vec\pi$ field, we have to
perform the rotations given by $\vartheta$ and
$\varphi$ on the states in the ensemble, which leads to a
spherically symmetric $\pi$ distribution at all proper times. 
%%%%%%%%%%%%%%%%%%%%%%
% initial submission %
%Therefore,
%for the neutral pion ratio \cite{rajwil93}, $R$, we obtain the
%distribution derived in ref.
%\cite{rajwil93}:
%%%%%%%%%%%%%%%%%%%%%%
% 1st revision       %
Therefore, 
even though unlike ref. \cite{rajwil93} we do {\em not} have 
an $O(4)$ isospin symmetry,
for the neutral pion ratio, $R$, 
we obtain exactly the same distribution:
%%%%%%%%%%%%%%%%%%%%%%
$$
  {\cal P}(R)=\frac{1}{2\sqrt{R}}.
$$

\subsection{Results for the ensemble of initial conditions}

In this subsection we shall present the results of the generated ensembles
of
initial conditions. These ensembles depend on the freeze-out temperature
and chemical potential, $T_0$ and $\mu_0$, and on the volume of the DCC
domain. The former two are determined since we start from a 20\%
supercooled
baryon-free 
plasma, i.e.,
$T_0=130$MeV and $\mu_0=0$.
%%%%%%%%%%%%%%%%%%%%%%
% initial submission %
%However, the volume of the domain is not
%so easy to estimate. 
%
%Hydrodynamical calculations predict a maximum of the plasma
%volume of about 50fm$^3$ \cite{csened94,Csernai94:Ch10:58}. How\-ever,
%hydrodynamical
%models assume local thermal equilibrium during the evolution, which is
%shown not to be the case in a real collision \cite{csocse94}.
%Thus, this value could be questionable. 
%
%Effective field theoretical calculations 
%in the linear sigma model \cite{huawan94,gavmul94,asahua95} 
%predict domain sizes of up to $3-5$fm (in the LR frame).
%This corresponds to a maximum  domain volume of about $25-125$fm$^3$. 
%
%We have chosen three different values for our simulations: $2$fm$^3$ to
%simulate
%the very small domains with a size slightly above 1fm, $10$fm$^3$ for
%domains with a size of about $2-2.5$fm, and $50$fm$^3$ for the really big
%domains of $3.5-4$fm size. The number of different initial conditions in
%each
%ensemble was 5000.
%%%%%%%%%%%%%%%%%%%%%%
% 1st revision       %

It is known \cite{rajwil93} that at $T\approx T_c$
the correlation length in the linear sigma model is comparable 
to the pion size. 
To simulate these most probable initial domain sizes, we have
chosen the value $V_0 = 0.5$fm$^3$.

To give a comparison to the Gaussian initial condition (\ref{gaussinit}),
we also tried $V_0=3.3$fm$^3$ since that gives exactly the same initial 
field derivative distributions.

Unfortunately, our model does not include the dynamical evolution of the
domain sizes. It is not clear what the survival rates for
domains of different sizes are and it may occur that relatively
large domains have the best chance to survive. That is why we chose a
third value, $V_0=10$fm$^3$, for our analysis.

All three initial ensembles contained 15,000 states. 
%%%%%%%%%%%%%%%%%%%%%%%%%

\subsubsection{Various distributions of the initial ensemble}
\label{Ssc:distens}

\begin{figure}[htbp]
\vspace*{-2cm}
\center
\leavevmode
%%%%%%%%%%%%%%%%%%%%%%
% initial submission %
% \epsfysize=18cm
%%%%%%%%%%%%%%%%%%%%%%
% 1st revision       %
\epsfysize=14.5cm
%%%%%%%%%%%%%%%%%%%%%%
\epsfbox{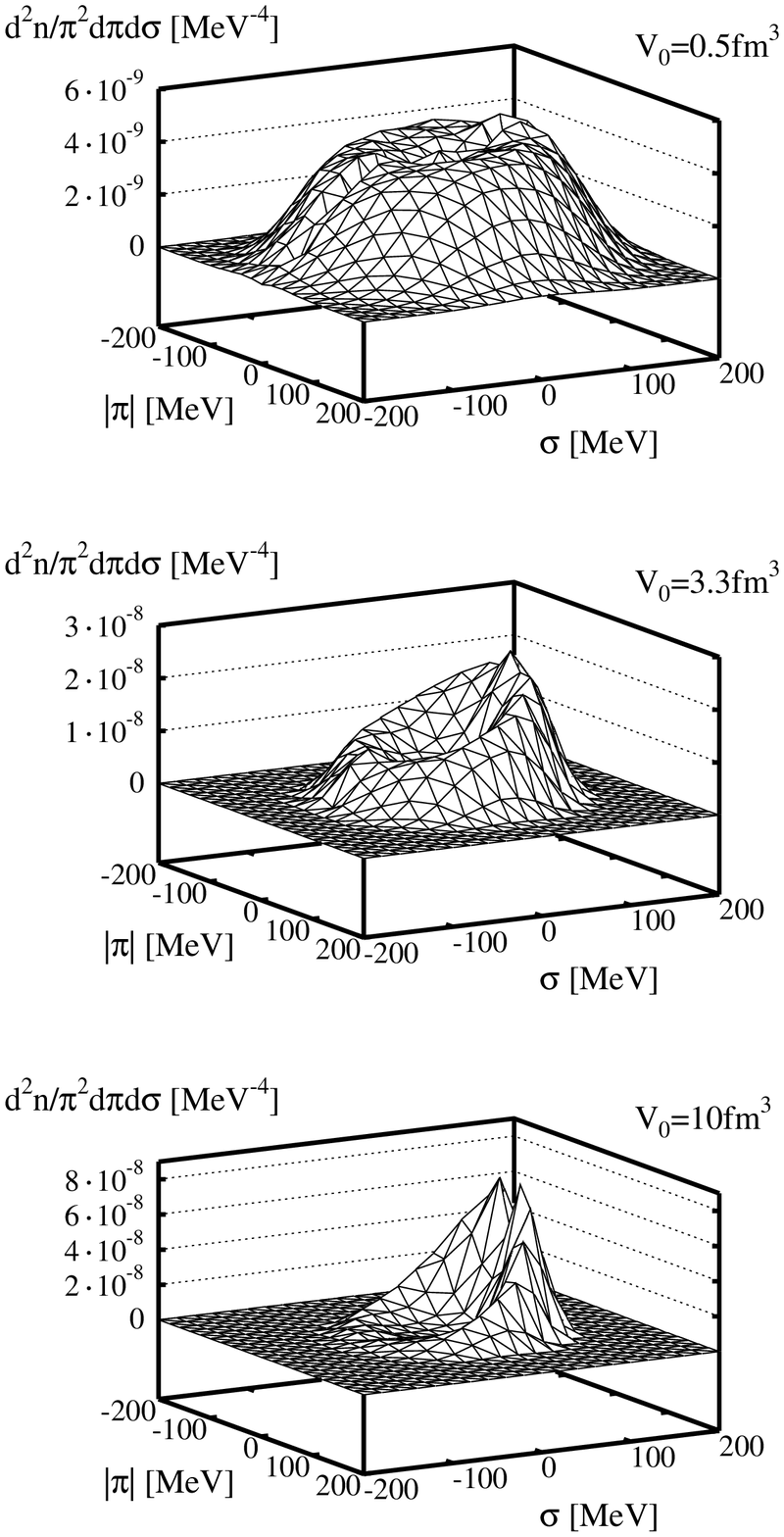}
%%%%%%%%%%%%%%%%%%%%%%
% initial submission %
%       \vspace*{-0.5cm}
%\caption[Initial distribution of the chiral fields $\sigma$ and $\pi_1$, 
%for initial 
%volumes $V_0=2$fm$^3$, 10fm$^3$, 50fm$^3$, at $T_0=130$MeV.] 
%%%%%%%%%%%%%%%%%%%%%%
% 1st revision       %
       \vspace*{0.3cm}
\caption[]
%%%%%%%%%%%%%%%%%%%%%%
{\footnotesize 
%%%%%%%%%%%%%%%%%%%%%%
% initial submission %
%The initial distribution of the chiral fields $\sigma$ and $\pi_1$, for
%initial 
%volumes $V_0=2$fm$^3$ (top), 10fm$^3$ (middle) and 50fm$^3$ (bottom),
%at the initial 
%temperature $T_0=130$MeV for an ensemble of 5000. 
%The distributions of 
%the other pion fields, 
%$\pi_2$ and $\pi_3$, are similar to that of $\pi_1$.  
%Manifestly, 
%the distribution is narrower for a larger initial volume.
%%%%%%%%%%%%%%%%%%%%%%
% 1st revision       %
The  initial distribution of the chiral field $\sigma$ 
and the magnitude of $\vec\pi$, 
for initial volumes 
$V_0=0.5$fm$^3$ (top), 3.3fm$^3$ (middle) and 10fm$^3$ (bottom), 
at the initial temperature $T_0=130$MeV for an ensemble of 600,000 states. 
Manifestly, 
the distribution has its maximum at a non-zero $\sigma$ value 
and becomes narrower for a larger initial volume.
To aid the eye, 
the distributions were complemented by a reflection 
on the $|\vec\pi|=0$ plane.
%%%%%%%%%%%%%%%%%%%%%%
}
       \label{fig:sp_2-10-50}
\end{figure}

\paragraph{The distribution of the chiral fields.}
Fig. \ref{fig:sp_2-10-50} shows the joint
distribution function\footnote{
%%%%%%%%%%%%%%%%%%%%%%
% initial submission %
%These distribution functions are obtained
%by introducing a rectangular grid in the $\sigma-\pi_1$ plane
%and counting the
%number of states in the ensemble in each grid cell. 
%The
%value of the distribution function in a given grid cell is approximated
%by the ratio of the number of states in the cell to the area of the cell.
%Since we
%have finite
%number of points in the ensemble, these plots depend
%on the size of the grid cells. For small cells statistical fluctuations
%are large, thus the plots are ragged, while a big cellsize smoothens out
%the
%distributions. We have chosen a cellsize of 15MeV in both the $\sigma$
%and
%$\pi_1$ direction.
%%%%%%%%%%%%%%%%%%%%%%
% 1st revision       %
These distributions were obtained
by introducing a rectangular grid in the $\sigma-|\vec\pi|$ plane
and counting the states in each grid cell 
with a $1/{\vec\pi}^2$ weight. 
The value of the distribution function in a given grid cell 
was approximated by the ratio of the total for the cell to the area 
of the cell.
Then we divided by the total number of states to normalize 
the distributions to unity.
Since the $1/{\vec\pi}^2$ Jacobian enhances statistical fluctuations 
for small $|\vec\pi|$ values tremendously,
we had to take a huge ensemble of 600,000 states for these plots.
These fluctuations also depend on the cellsize; 
the distributions are smoother for bigger cells.
We have chosen a cellsize of 20MeV in 
both the $\sigma$ and $|\vec\pi|$ direction.
%%%%%%%%%%%%%%%%%%%%%%
}
%%%%%%%%%%%%%%%%%%%%%%
% initial submission % 
%of the chiral fields $\sigma$ and $\pi_1$
%in
%the initial ensemble. The $\sigma-\pi_2$ and $\sigma-\pi_3$ distributions
%are similar, which is a direct consequence of the rotational
%symmetry discussed in the previous subsection.
%As we expected,
%the peak of the distribution is near to the minimum of the potential
%energy density
%Thus, the realistic initial conditions are around this minimum, 
%which is at $(\sigma,\vec\pi)=(40{\rm
%MeV},\vec 0)$ at the temperature $T_0=130$MeV. Hence, the initial
%condition (\ref{usinit}), which assumes $(\sigma,\vec\pi)=(0,\vec 0)$,
%is rather unlikely.
%%%%%%%%%%%%%%%%%%%%%%
% 1st revision       %
of the chiral field $\sigma$ and the length of the isovector $\vec\pi$ 
in the initial ensemble.
As we expected,
the peak of the distribution is not at $\sigma = 0$, 
however it is not at the minimum ($\sigma\approx 40$MeV) 
of the potential energy density, either. 
This is because of the quarks.
The quark energy density is a monotone decreasing function
of $\sigma^2+\vec\pi^2$, therefore  it is more favourable energetically
to have a larger condensate, $\approx 86$MeV.
%%%%%%%%%%%%%%%%%%%%%%

For bigger volumes the distributions are
sharper. The reason is that
fluctuations are smaller in larger systems. Thus, the probability for the
system to be far from the minimum of the potential energy density is
larger for small systems. This increases the probability of having 
large oscillations, which helps the observability of the domain.
However, this increase may be
compensated
by the fact that small domains emit fewer pions. 
%%%%%%%%%%%%%%%%%%%%%%%%%%%
% 1st revision - inserted %
Unfortunately, our one-dimensional model does not allow us to decide this
question 
because it includes only a scaling expansion of the domains 
but cannot account for domain evolution due to instabilities. 
%%%%%%%%%%%%%%%%%%%%%%%%%%%

As a comparison, 
Fig. \ref{fig:birmol97:fig7} shows 
%%%%%%%%%%%%%%%%%%%%%%
% initial submission %
%the $\sigma-\pi_1$ distribution 
%%%%%%%%%%%%%%%%%%%%%%
% 1st revision       %
the $\sigma-|\vec\pi|$ distribution
%%%%%%%%%%%%%%%%%%%%%%
for the
microcanonical ensemble shown in Fig. 7 in ref. \cite{birmol97}. The
states in the ensemble 
%%%%%%%%%%%%%%%%%%%%%%
% initial submission %
%are concentrated around two circles, 
%%%%%%%%%%%%%%%%%%%%%%
% 1st revision       %
are concentrated in a circular band,
%%%%%%%%%%%%%%%%%%%%%% 
which 
is rather different from our peaked distributions. This is a
consequence of the 
%%%%%%%%%%%%%%%%%%%%%%
% initial submission %
%very high 
%%%%%%%%%%%%%%%%%%%%%%
% 1st revision       %
high
%%%%%%%%%%%%%%%%%%%%%%
initial chiral "angular momentum" and the
constraint that these initial states all have the same energy.

\begin{figure}[htbp]
\vspace*{-0.5cm}
\center
\leavevmode
\epsfysize=6cm
\epsfbox{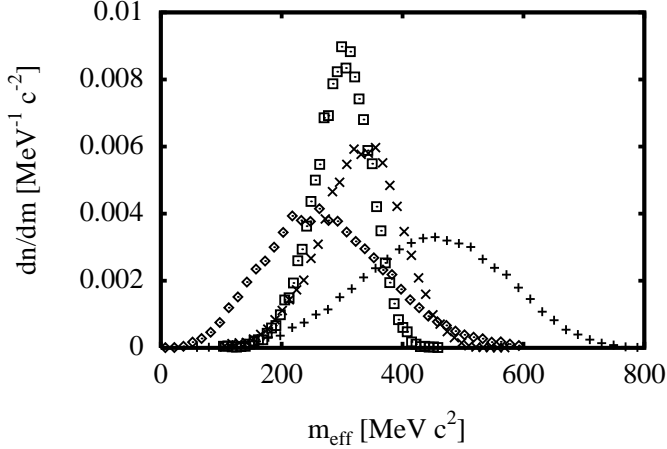}
       \vspace*{0cm}
%%%%%%%%%%%%%%%%%%%%%%
% initial submission %
%\caption[Initial distribution of the effective quark/anti\-quark mass,
%for 
%initial 
%volumes $V_0=2$fm$^3$, 10fm$^3$, 50fm$^3$, at $T_0=130$MeV.] 
%%%%%%%%%%%%%%%%%%%%%%
% 1st revision       %
\caption[]
%%%%%%%%%%%%%%%%%%%%%%
{\footnotesize
The initial distribution of the effective quark/antiquark mass 
for 
%%%%%%%%%%%%%%%%%%%%%%
% initial submission %
%initial volumes\ \  $V_0=\ 2$fm$^3$ (dotted line),\ \ 
%10fm$^3$ (dashed line) and 
%50fm$^3$ (solid line),\ \  
%at the initial temperature\ \  $T_0=130$MeV for an ensemble of 5000. 
%%%%%%%%%%%%%%%%%%%%%%
% 1st revision       %
initial volumes $V_0=0.5$fm$^3$ (pluses), 3.3fm$^3$ (crosses) 
and 10fm$^3$ (boxes),  
at the initial temperature $T_0=130$MeV for an ensemble 
of 15,000 states. 
The 'diamonds' correspond 
to the Gaussian initial condition (\ref{gaussinit}).
%%%%%%%%%%%%%%%%%%%%%%
As the initial volume decreases, mass fluctuations intensify and the
average 
mass increases.
}
       \label{fig:m_2-10-50}
\end{figure}

\paragraph{The distribution of the quark/antiquark effective mass.}
%%%%%%%%%%%%%%%%%%%%%%
% initial submission %
%A consequence of the initial distribution of the chiral fields is the
%fact that 
%%%%%%%%%%%%%%%%%%%%%%
% 1st revision       %
The initial distribution of the chiral fields implies that
%%%%%%%%%%%%%%%%%%%%%%
the initial quark mass is very unlikely to be zero, 
which again questions the appropriateness of the choice (\ref{usinit}). 
The actual distribution\footnote{
All one-dimensional distributions were calculated 
by dividing the plotted interval into $40-50$ equal sections
and counting the number of states in each section.
%%%%%%%%%%%%%%%%%%%%%%%%%%%
% 1st revision - inserted %
All these distributions are normalized to unity.
%%%%%%%%%%%%%%%%%%%%%%%%%%%
} 
of the effective quark mass is shown in Fig.
\ref{fig:m_2-10-50}. 
As the initial volume decreases, the distribution broadens and the average
quark/antiquark mass increases. 

\paragraph{The distribution of the proper time derivatives of the fields.}
By Eqs. 
(\ref{candist}) and (\ref{endens}),
the proper time derivatives of the chiral fields follow Gaussian
distributions with
the same width, $\sqrt{T_0/V_0}$.
For a given freeze-out temperature,
the width 
is {\em inversely} proportional to $\sqrt{V_0}$.
In ref. \cite{rajwil93_2} and followers
\cite{gavgoc94,huawan94,biaczy95,asahua95} this
width is
assumed to be order of $v$, interpreted as $v$/1fm, 
which is
$88{\rm MeV/fm}$. This is the same order of magnitude as
our results and would correspond to $V_0=3.3{\rm fm}^3$ at $T_0=130$MeV.
To obtain a significantly lower width, one has to take an
extremely
low freeze-out temperature and/or a very large domain size.

\begin{figure}[htbp]
\vspace*{-0.5cm}
\center
\leavevmode
\epsfysize=6cm
\epsfbox{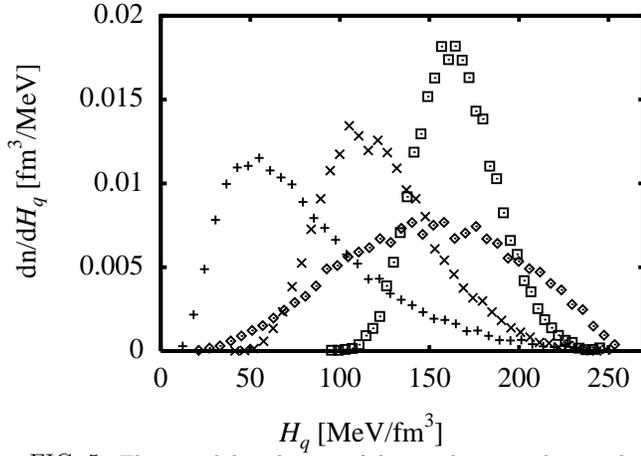}
       \vspace*{0cm}
%%%%%%%%%%%%%%%%%%%%%%
% initial submission %
%\caption[Initial distribution of the quark energy density (including 
%antiquarks), for initial 
%volumes $V_0=2$fm$^3$, 10fm$^3$, 50fm$^3$, at $T_0=130$MeV.] 
%%%%%%%%%%%%%%%%%%%%%%
% 1st revision       %
\caption[]
%%%%%%%%%%%%%%%%%%%%%%
{\footnotesize
The initial distribution of the quark energy density
density (\protect\ref{endens}) for 
%%%%%%%%%%%%%%%%%%%%%%
% initial submission %
%initial volumes $V_0=2$fm$^3$ (dotted line), 10fm$^3$  (dashed line)
%and 50fm$^3$ (solid line), 
%at the initial temperature $T_0=130$MeV for an ensemble of 5000. 
%%%%%%%%%%%%%%%%%%%%%%
% 1st revision       %
initial volumes $V_0=0.5$fm$^3$ (pluses), 3.3fm$^3$  (crosses)
and 10fm$^3$ (boxes), 
at the initial temperature $T_0=130$MeV for an ensemble of 15,000 states. 
The 'diamonds' correspond 
to the Gaussian initial condition (\ref{gaussinit}).
%%%%%%%%%%%%%%%%%%%%%%
Since for smaller initial 
volumes the quark mass tends to be bigger, the quark energy density is 
then smaller. }
  \label{fig:qe_2-10-50}
\end{figure}

\begin{figure}[htbp]
\vspace*{-0.8cm}
\center
\leavevmode
\epsfysize=6cm
\epsfbox{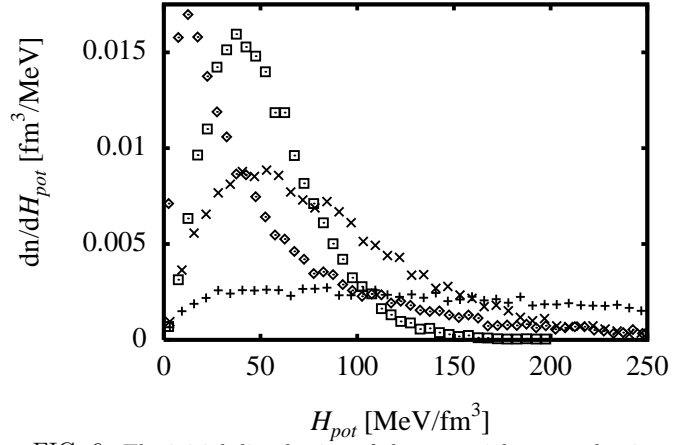}
       \vspace*{0cm}
%%%%%%%%%%%%%%%%%%%%%%
% initial submission %
%\caption[Initial distribution of the potential 
%energy density of the meson fields, for initial 
%volumes $V_0=2$fm$^3$, 10fm$^3$, 50fm$^3$, at $T_0=130$MeV.] 
%%%%%%%%%%%%%%%%%%%%%%
% 1st revision       %
\caption[]
%%%%%%%%%%%%%%%%%%%%%%
{\footnotesize
The initial distribution of the potential energy 
density (\protect\ref{endens}) of the meson fields for
%%%%%%%%%%%%%%%%%%%%%%
% initial submission %
%initial volumes $V_0=2$fm$^3$ (dotted line), 10fm$^3$ (dashed line) and 
%50fm$^3$ (solid line), 
%at the initial temperature $T_0=130$MeV for an ensemble of 5000. 
%%%%%%%%%%%%%%%%%%%%%%
% 1st revision       %
initial volumes $V_0=0.5$fm$^3$ (pluses), 3.3fm$^3$ (crosses) 
and 10fm$^3$ (boxes),
at the initial temperature $T_0=130$MeV for an ensemble of 15,000 states. 
The 'diamonds' correspond 
to the Gaussian initial condition (\ref{gaussinit}).
%%%%%%%%%%%%%%%%%%%%%%
As the initial volume decreases, the system is more and more probable 
to be on the walls of the valley of the potential energy density, 
far from the minimum. }
  \label{fig:pe_2-10-50}
\end{figure}

\begin{figure}[htbp]
%%%%%%%%%%%%%%%%%%%%%%
% initial submission %
%\vspace*{-0.5cm}
%%%%%%%%%%%%%%%%%%%%%%
% 1st revision       %
\vspace*{-1.5cm}
%%%%%%%%%%%%%%%%%%%%%%
\center
\leavevmode
\epsfysize=6cm
\epsfbox{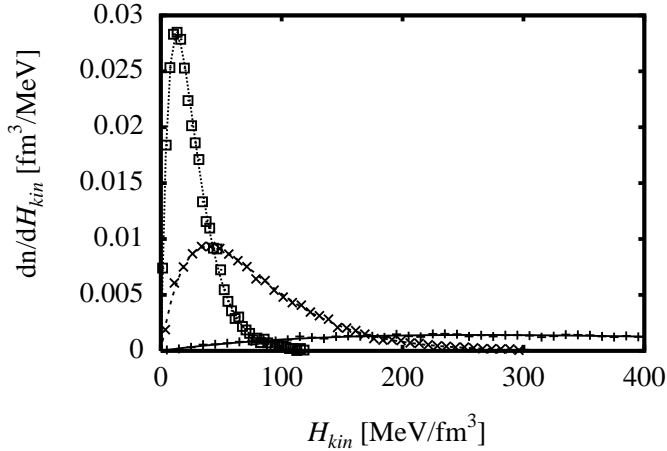}
       \vspace*{0cm}
%%%%%%%%%%%%%%%%%%%%%%
% initial submission %
%\caption[Initial distribution of the 
%kinetic energy density of the meson fields, for initial 
%volumes $V_0=2$fm$^3$, 10fm$^3$, 50fm$^3$, at $T_0=130$MeV.] 
%%%%%%%%%%%%%%%%%%%%%%
% 1st revision       %
\caption[]
%%%%%%%%%%%%%%%%%%%%%%
{\footnotesize
The initial distribution of the kinetic energy 
density (\protect\ref{endens}) of the meson fields for
%%%%%%%%%%%%%%%%%%%%%%
% initial submission %
%initial volumes $V_0=2$fm$^3$ (dotted line), 10fm$^3$ (dashed line)
% and 
%50fm$^3$ (solid line), 
%at the initial 
%temperature $T_0=130$MeV for an ensemble of 5000. 
%%%%%%%%%%%%%%%%%%%%%%
% 1st revision       %
initial volumes $V_0=0.5$fm$^3$ (pluses), 3.3fm$^3$ (crosses) 
and 10fm$^3$ (boxes), 
at the initial temperature $T_0=130$MeV for an ensemble of 15,000 states. 
The distribution for the Gaussian initial condition (\ref{gaussinit}) 
is identical to taking $V_0=3.3$fm$^3$.
The agreement with the theoretical curves is excellent.
%%%%%%%%%%%%%%%%%%%%%%
For smaller 
initial volumes fluctuations are larger, and thus the kinetic energy 
distribution is broader and the average kinetic energy density is larger.}
       \label{fig:ke_2-10-50}
\end{figure}

\paragraph{The distribution of the quark, potential and kinetic energy
densities.}
The distribution of the quark, potential and kinetic energy densities  of 
the initial conditions 
can be seen in Figs. \ref{fig:qe_2-10-50}, \ref{fig:pe_2-10-50}, 
and \ref{fig:ke_2-10-50}. 
%%%%%%%%%%%%%%%%%%%%%%
% initial submission %
%We can see that quarks carry only a small fraction of the energy density,
%while the kinetic and potential energy density have the same order of
%magnitude. 
%%%%%%%%%%%%%%%%%%%%%%
% 1st revision       %
We can see that the kinetic, potential and quark energy densities 
have  the same order of magnitude.
%%%%%%%%%%%%%%%%%%%%%%

Due to the increasing fluctuations,\ \  all distributions spread
as the initial volume decreases. Furthermore, the peaks of the kinetic and
potential
energy density distributions shift to higher values. This is in 
agreement with our picture: more fluctuations result in more kinetic
energy density, and thus the system climbs higher and higher on the wall
of the valley of the potential energy density. On the other hand, for
the quarks the peak
shifts to lower values for increasing initial volumes. This is due to the
fact that
fluctuations 
%%%%%%%%%%%%%%%%%%%%%%%%%%%%%%%%%%%%%%%%%%%%%%%%
% initial submission - deleted in 1st revision %
%tend to 
%%%%%%%%%%%%%%%%%%%%%%%%%%%%%%%%%%%%%%%%%%%%%%%%
increase the quark/antiquark effective mass. 

The distribution of the kinetic energy density (\ref{endens}) can be
calculated
analytically. 
%%%%%%%%%%%%%%%%%%%%%%
% initial submission %
%We know that the distributions of the field derivatives are
%Gaussians:
%\begin{eqnarray}
%  p(\dot{\vec\Phi})\,d\dot{\vec\Phi}&\propto&  
% \exp\left(-\frac{V_0}{2T_0}\dot\Phi^2\right)
%d\dot{\vec\Phi} \nonumber\\
%&\propto& \exp\left(-\frac{V_0}{2T_0}\dot\Phi^2\right)
%\dot\Phi^3
%\, d\dot\Phi\, d\Omega_{4D}
%\ ,
%\end{eqnarray}
%where $\Omega_{4D}$ is the 4-dimensional solid angle element.
%Thus, since ${\cal H}_{kin}=\dot{\vec\Phi}^2/2$, we obtain
%$$
%  f({\cal H}_{kin})\,d{\cal H}_{kin}\propto
%\exp\left(-\frac{V_0}{T_0}{\cal
%H}_{kin}\right)
%{\cal
%H}_{kin}\, d{\cal H}_{kin} \ .
%$$
%%%%%%%%%%%%%%%%%%%%%%
% 1st revision       %
Using that ${\cal H}_{kin}=\dot{\vec\Phi}^2/2$,
from Eq. (\ref{candistpolar}) 
it is not difficult to obtain that
$$
  f({\cal H}_{kin})\,d{\cal H}_{kin} = \left(\frac{V_0}{T_0}\right)^2
\exp\left(-\frac{V_0}{T_0}{\cal
H}_{kin}\right)
{\cal
H}_{kin}\, d{\cal H}_{kin} \ .
$$
%%%%%%%%%%%%%%%%%%%%%%
A simple calculation yields that the 
maximum of the initial kinetic energy
distribution 
is at ${\cal H}_{kin}=T_0/V_0$. This value is $2-3$ orders of magnitude
larger than the kinetic energy
density corresponding to the initial condition (\ref{usinit}), which is
about $0.066{\rm MeV/fm^3}$.

\subsubsection{The evolution of the ensemble}

\begin{figure}[htbp]
\vspace*{-0.5cm}
\center
\leavevmode
%%%%%%%%%%%%%%%%%%%%%%
% initial submission %
%\epsfysize=11.5cm
%\hspace*{-.7cm}
%\epsfbox{fig8.eps}
%       \vspace*{-0.5cm}
%\caption[Time evolution of the average and standard deviation of
%the $\sigma$ and $\pi_1$ fields for initial
%volumes $V_0=2$fm$^3$, 10fm$^3$, 50fm$^3$, 
%with $\tau_0=1$fm/$c$ and $T_0=130$MeV.] 
%%%%%%%%%%%%%%%%%%%%%%
% 1st revision       %
\epsfysize=9.0cm
\epsfbox{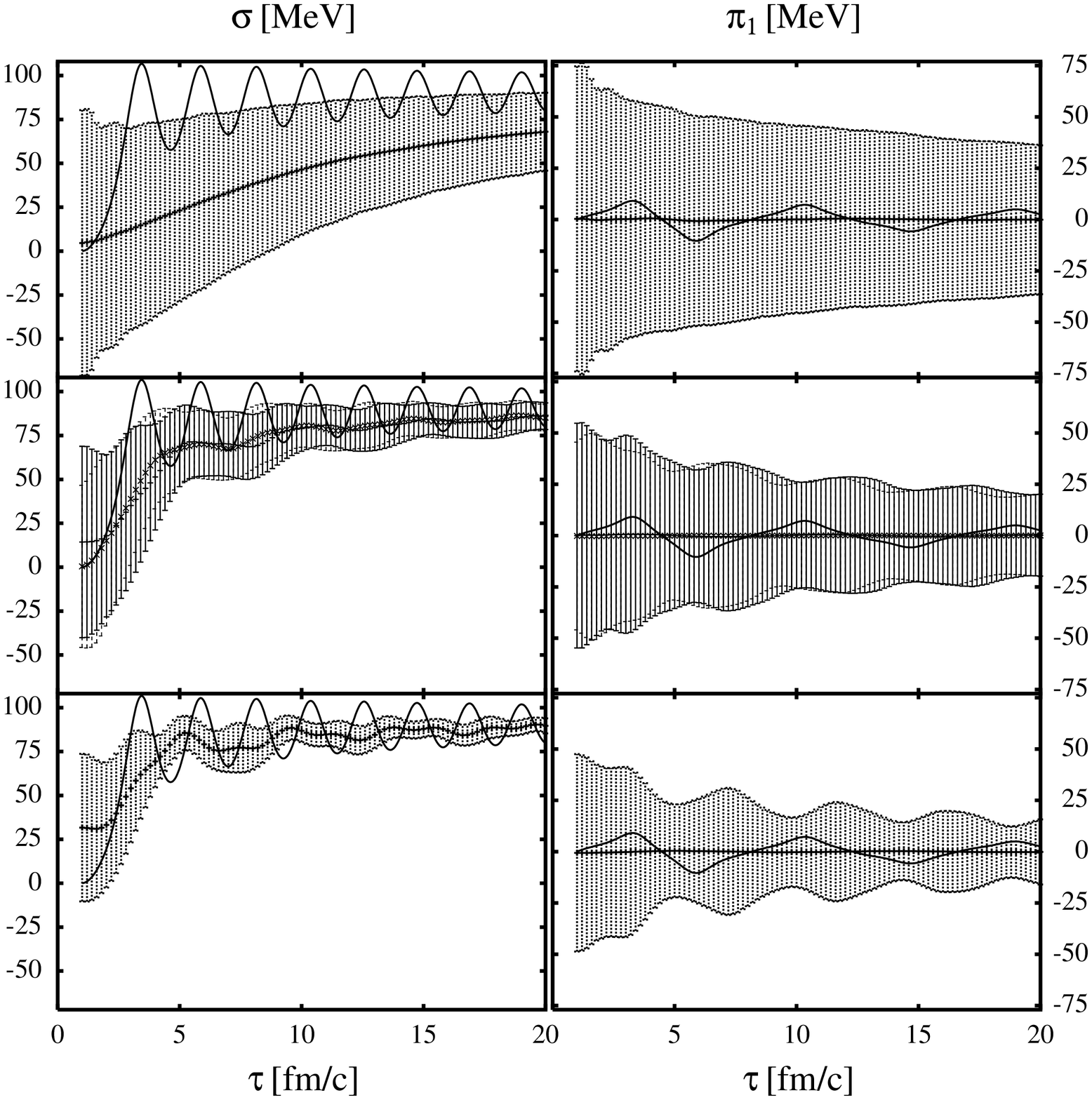}
       \vspace*{0.3cm}
\caption[]
%%%%%%%%%%%%%%%%%%%%%%
{\footnotesize
%%%%%%%%%%%%%%%%%%%%%%
% initial submission %
%The time evolution of the average and standard deviation of the $\sigma$
%and $\pi_1$
%fields for 
%initial volumes $V_0=2$fm$^3$ (top), 10fm$^3$ (middle) and 
%50fm$^3$ (bottom), 
%with a freeze-out time $\tau_0=1$fm/$c$ at the initial 
%temperature $T_0=130$MeV, for an ensemble of 5000. 
%The dashed regions correspond to one standard deviation from the mean (at 
%least 60\% confidence). 
%The thin solid lines correspond to the evolution starting from
%the initial condition
%(\protect\ref{usinit}). 
%%%%%%%%%%%%%%%%%%%%%%
% 1st revision       %
The time evolution of the ensemble average and "70\% interval" 
of the $\sigma$ and $\pi_1$ fields
for initial volumes $V_0=0.5$fm$^3$ (top), 3.3fm$^3$ (middle) 
and 10fm$^3$ (bottom), 
with a freeze-out time $\tau_0=1$fm/$c$,
initial temperature $T_0=130$MeV, 
for an ensemble of 15,000. 
The shaded regions represent the 70\% fraction centered 
on the mean value. 
The thin solid lines correspond to the evolution starting 
from the initial condition (\protect\ref{usinit}).
%%%%%%%%%%%%%%%%%%%%%%
The evolution of the $\sigma$ field is rather uncharacteristic for the
majority of
the ensemble, showing that this initial condition is improbable. However,
the oscillations of the pion fields are well within 
range, which is due to the rather small kinetic energy corresponding to 
(\protect\ref{usinit}).
%%%%%%%%%%%%%%%%%%%%%%%%%%%
% 1st revision - inserted %
The middle figures also show the ensemble evolution 
for the Gaussian initial condition (\protect\ref{gaussinit}); 
mean values -- crosses, 70\% interval -- dashed. 
The evolution is close to that from $V_0=3.3$fm$^3$.
%%%%%%%%%%%%%%%%%%%%%%%%%%%
}
       \label{fig:e_si-pi1_1}
\end{figure}

\begin{figure}[htbp]
\vspace*{-0.5cm}
\center
\leavevmode
%%%%%%%%%%%%%%%%%%%%%%
% initial submission %
%\epsfysize=12.0cm
%\hspace*{-0.7cm}
%\epsfbox{fig9.eps}
%       \vspace*{-0.5cm}
%\caption[Time evolution of the average and standard deviation of
%the $\sigma$ and $\pi_1$ fields for initial 
%volumes $V_0=2$fm$^3$, 10fm$^3$, 50fm$^3$, 
%with $\tau_0=7$fm/$c$ and $T_0=130$MeV.] 
%%%%%%%%%%%%%%%%%%%%%%
% 1st revision       %
\epsfysize=9.0cm
\epsfbox{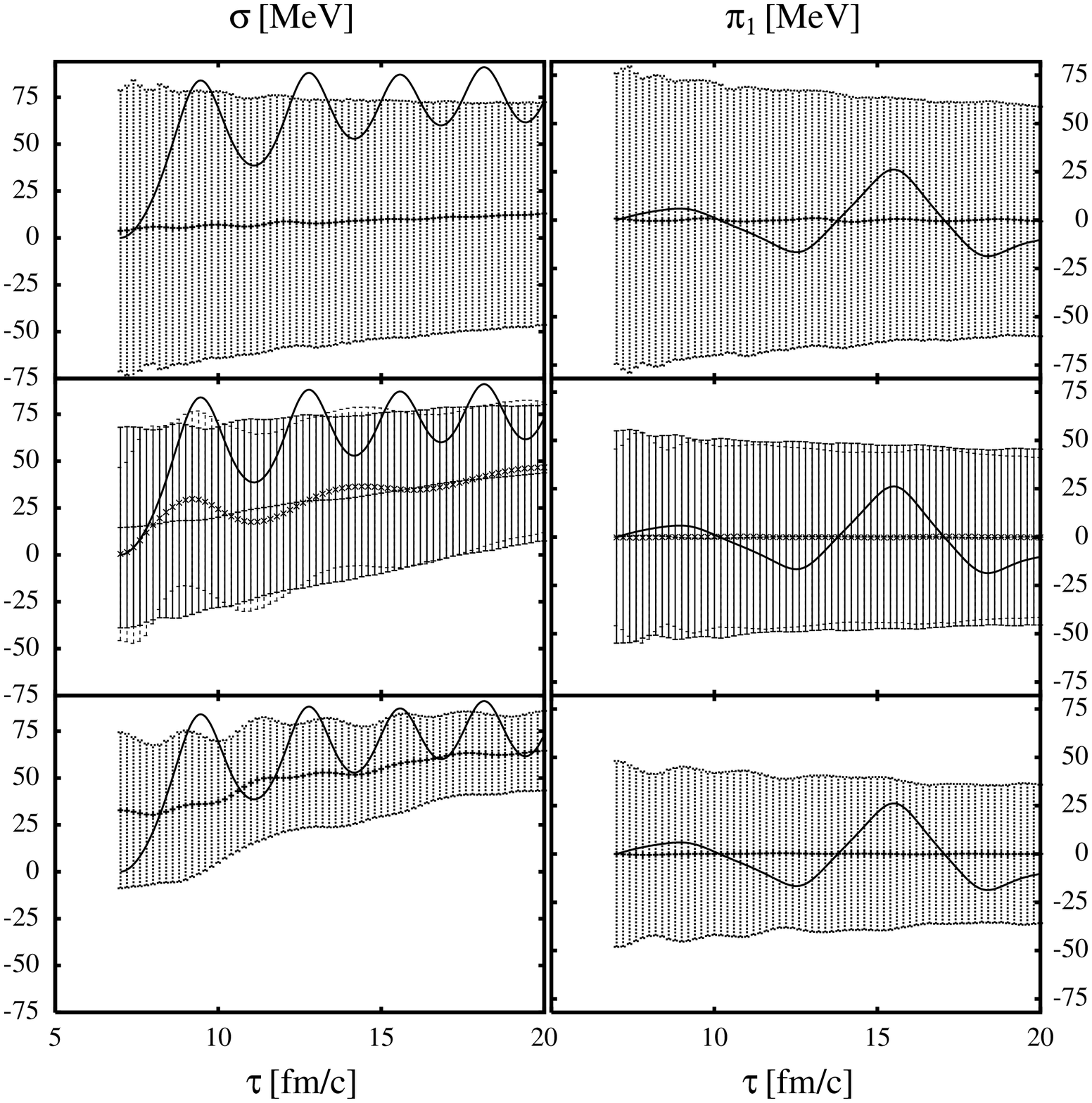}
       \vspace*{0.3cm}
\caption[]
%%%%%%%%%%%%%%%%%%%%%%
{\footnotesize
%%%%%%%%%%%%%%%%%%%%%%
% initial submission %
%The time evolution of the average and standard deviation of the $\sigma$ 
%(left column) and $\pi_1$ (right column)
%fields for
%initial volumes $V_0=2$fm$^3$ (top), 10fm$^3$ (middle) and 
%50fm$^3$ (bottom), 
%with a freeze-out time $\tau_0=7$fm/$c$ at the initial 
%temperature $T_0=130$MeV, for an ensemble of 5000. 
%The shaded areas correspond to one standard deviation from the mean value 
%(at least 60\% confidence). 
%%%%%%%%%%%%%%%%%%%%%%
% 1st revision       %
The time evolution of the ensemble average and "70\% interval" 
of the $\sigma$ and $\pi_1$ fields
for initial volumes $V_0=0.5$fm$^3$ (top), 3.3fm$^3$ (middle) 
and 10fm$^3$ (bottom), 
with a freeze-out time $\tau_0=1$fm/$c$,
initial temperature $T_0=130$MeV, 
for an ensemble of 15,000. 
The shaded regions represent the 70\% fraction centered 
on the mean value. 
The thin solid lines correspond to the evolution starting 
from the initial condition (\protect\ref{usinit}).
%%%%%%%%%%%%%%%%%%%%%%
The evolution of the $\sigma$ field starting 
from the initial 
condition (\protect\ref{usinit})  is still
uncharacteristic 
for the majority of
the ensemble. 
%%%%%%%%%%%%%%%%%%%%%%
% initial submission %
%Furthermore, for the biggest domains even the pion fields
%can 
%leave the confidence region. 
%%%%%%%%%%%%%%%%%%%%%%
% 1st revision       %
However, 
the oscillations of the pion fields are well within range, 
which is due to the rather small kinetic energy 
corresponding  to (\protect\ref{usinit}).
The bottom figures also show the ensemble evolution 
for the Gaussian initial condition (\protect\ref{gaussinit}); 
mean values -- crosses, 70\% interval -- dashed. 
The evolution is close to that from $V_0=3.3$fm$^3$.
%%%%%%%%%%%%%%%%%%%%%%
}
       \label{fig:e_si-pi1_7}
\end{figure}

\paragraph{The evolution of the ensemble of the chiral fields.}
Figs. \ref{fig:e_si-pi1_1} and \ref{fig:e_si-pi1_7} show the time 
evolution of 
%%%%%%%%%%%%%%%%%%%%%%
% initial submission %
%the average and standard deviation 
%%%%%%%%%%%%%%%%%%%%%%
% 1st revision       %
the ensemble average and the 70\% fraction
%%%%%%%%%%%%%%%%%%%%%%
of the chiral fields, $\sigma$ and $\pi_1$, for
freeze-out
times $\tau_0=1$fm/$c$ and 7fm/$c$. The other two pion
fields, $\pi_2$ and $\pi_3$, have a similar evolution to that of $\pi_1$
because of the rotational symmetry. 

%%%%%%%%%%%%%%%%%%%%%%
% initial submission %
%Here, by average and standard deviation we mean the arithmetic mean,
%$m_A$,
%$$
%  m_A=\frac{1}{N}\sum\limits_{i=1}^N A_i \ ,
%$$
%and the empirical standard deviation, $\sigma_A$, 
%$$
%  \sigma_A^2=\frac{1}{N-1}\sum\limits_{i=1}^N \left(A_i-m_A\right)^2
%\ ;
%$$
%where $A_i$ denotes the elements in the representative ensemble of the
%quantity $A$. 
%The confidence level  
%for one and two standard deviations from the mean, i.e., the fraction of
%the 
%ensemble in the corresponding intervals, was checked and was found to
%be at 
%least 60\% and 90\%, respectively.
%%%%%%%%%%%%%%%%%%%%%%%
% 1st revision        %
Here, by average we mean the arithmetic mean,
while the "70\% fraction" is the interval 
that is centered on the average and contains 70\% of the ensemble. 
One could also use the usual estimate of the standard deviation but,
because the distributions are non-Gaussian, 
that would not correspond to the same fraction of the ensemble 
at different proper times 
(we determined that in our case one standard deviation means 
an at least 60\% fraction, 
while two standard deviations correspond to at least 90\%).
%%%%%%%%%%%%%%%%%%%%%%%

%%%%%%%%%%%%%%%%%%%%%%
% initial submission %
%The confidence regions help us to
%%%%%%%%%%%%%%%%%%%%%%
% 1st revision       %
The 70\% fraction helps us
%%%%%%%%%%%%%%%%%%%%%%
pick out very unlucky events, i.e., 
those which are very far out of 
%%%%%%%%%%%%%%%%%%%%%%
% initial submission %
%the confidence region.
%The confidence 
%regions 
%%%%%%%%%%%%%%%%%%%%%%
% 1st revision       %
the interval.
These intervals
%%%%%%%%%%%%%%%%%%%%%%
are narrower for bigger systems, due to the well-known fact that 
fluctuations are smaller in bigger systems \cite{Huang66,Reif65,LanLif5}. 

The time evolution from the initial condition
(\ref{usinit}) with $\tau_0=1$fm/$c$ is rather uncharacteristic 
for the majority, at least 90\%, 
of
the ensemble, 
%%%%%%%%%%%%%%%%%%%%%%
% initial submission %
%as the $\sigma$ field is most of the time away from the
%average by 
%more than two standard deviations. 
%%%%%%%%%%%%%%%%%%%%%%
% 1st revision       %
as from time to time the $\sigma$ field is away from the
average by 
about two standard deviations. 
For $\tau_0=7$fm/$c$ the situation is slightly better 
- the above fraction is 80\%.  
This is because the damping is reduced due to the slower expansion,
therefore the distribution becomes wider.
We conclude, 
that the initial condition (\ref{usinit}) is representative only 
if the initial ensemble has larger fluctuations, i.e., 
we have a much higher initial temperature or smaller initial volume,
which is physically unreasonable.
%%%%%%%%%%%%%%%%%%%%%%%%

However, the evolution of
the pion fields is well in 
%%%%%%%%%%%%%%%%%%%%%%
% initial submission %
%the 60\% confidence region. 
%%%%%%%%%%%%%%%%%%%%%%
% 1st revision       %
the 70\% fraction.
%%%%%%%%%%%%%%%%%%%%%%
Thus, the initial condition (\ref{usinit})
does {\em not}
overestimate 
pion oscillations.
Furthermore, due to the small starting kinetic energy, the 
oscillations are deep in the 
%%%%%%%%%%%%%%%%%%%%%%
% initial submission %
%confidence region, 
%%%%%%%%%%%%%%%%%%%%%%
% 1st revision       %
70\% interval,
%%%%%%%%%%%%%%%%%%%%%%
showing that oscillations with larger amplitudes would
be
similarly 
probable.
%%%%%%%%%%%%%%%%%%%%%%%%%%%
% 1st revision - inserted % 
The oscillations are out of the 70\% fraction only 
if the initial ensemble contains little fluctuations, i.e., 
if the initial temperature is very low and/or the initial volume 
is very large. 

Figs. (\ref{fig:e_si-pi1_1}) and (\ref{fig:e_si-pi1_7}) also tell 
that the Gaussian initial condition (\ref{gaussinit}) results 
in a very similar ensemble evolution 
to that for $V_0=3.3$fm$^3$ and $T_0=130$MeV. 
The 70\% intervals are almost identical, 
showing that both ensembles are concentrated in the same regions. 
The mean values are identically zero for the $\pi_1$ distribution 
as a consequence of the rotational symmetry. 
However, 
the $\sigma$ means are quite different 
because the distributions have different shapes 
and the initial distribution (\ref{gaussinit}) is peaked at $\sigma = 0$,
which is not the correct value of the condensate at $T_0=130$MeV.
%%%%%%%%%%%%%%%%%%%%%%%%%%%

\section{Conclusions}

In this work we investigated chiral condensate evolution
starting from a
canonical ensemble of initial states. In the scenario we presented, the
initial ensemble and its subsequent evolution is
characterized by three simple parameters: the freeze-out temperature,
$T_0$, the initial volume, $V_0$, and the freeze-out time $\tau_0$. 

We compared our initial condition to the "roll-down" scenario,
i.e., to starting from the unstable maximum of the potential with a little
kick \cite{huawan94,S96,fenmol97}. We found that the sigma field evolution
in the
"roll-down"
scenario is unrealistic because the sigma field oscillates counterphase,
however 
pion oscillations are not overestimated if we choose reasonable values for
the parameters, i.e., if  $T_0=130$MeV, 
%%%%%%%%%%%%%%%%%%%%%%
% initial submission %
%$V_0 \sim 2-50$fm$^3$,
%%%%%%%%%%%%%%%%%%%%%%
% 1st revision       %
$V_0 \sim 0.5-10$fm$^3$,
%%%%%%%%%%%%%%%%%%%%%%
and $\tau_0 \sim 1-7$fm/$c$.
%%%%%%%%%%%%%%%%%%%%%%%%%%%
% 1st revision - inserted %
Thus we confirm the suggestions \cite{blakrz96} 
and the findings \cite{birgre97} that observable DCC is a rare event 
in a canonical ensemble.
%%%%%%%%%%%%%%%%%%%%%%%%%%%
 
We also made a comparison to the often-quoted Gaussian field and field
derivative distributions with widths $v$ and $v/1$fm respectively
\cite{rajwil93_2,gavgoc94,huawan94,biaczy95}.
We found that these distributions are 
%%%%%%%%%%%%%%%%%%%%%%
% initial submission %
%nearly identical
%%%%%%%%%%%%%%%%%%%%%%
% 1st revision       %
quite close
%%%%%%%%%%%%%%%%%%%%%%
to the choice
$T_0=130$MeV and $V_0=3.3$fm$^3$. However, this choice also accounts for
the necessary non-zero mean of the $\sigma$ distribution corresponding to
the non-zero condensate at a finite temperature, $T_0 \sim 130$MeV.

We believe that our simple scenario gives more insight on what the
proper
initial
conditions could be. 
Though our results are similar to those obtained from other initial
conditions, the ones mentioned above and their modifications
\cite{asahua95},
we did contribute to reducing the
arbitrariness in those.
Our three parameters have clear physical meaning,
and they give a simple way and a good handle on how to choose the initial
field distributions.

One should keep in mind that all our calculations were done in a
one-dimensional Bjorken scaling picture. To obtain quantitative answers on
domain
sizes one inevitably has to take into account the spatial dimensions as
well. Our test calculations based on the linear response
method indicate that the fastest growth rates are about $1c$/fm,
thus the biggest 
domains are $3-4$fm large. This is in agreement with earlier
findings \cite{gavgoc94,huawan94,asahua95,csemis95,S96}.
However, we would like to emphasize that the only way to make real
quantitative statements about the observability of DCCs is to go one step
further and calculate observables,
%%%%%%%%%%%%%%%%%%%%%%
% initial submission %
%i.e., 
%%%%%%%%%%%%%%%%%%%%%%
% 1st revision       %
e.g.,
%%%%%%%%%%%%%%%%%%%%%%
to obtain the
spectra of pions emitted by the domains.

\section*{Acknowledgements}
Enlightening discussions with T. Bir\'o and M. Gyulassy are
gratefully
acknowledged.
This work was supported by the Research Council of Norway (through
its programs for nuclear and particle physics, supercomputing and free
projects), and by the Director, Office of Energy Research, Division 
of Nuclear Physics of the Office of High Energy and Nuclear Physics of
the  US Department of Energy under Contract Number DE-FG02-93ER40764.

%%%%%%%%%%%%%%% BIBLIOGRAPHY %%%%%%%%%%%%

%%%%%%%%%%%%%%%% FIGURES %%%%%%%%%%%%%%%%%%

\newpage

% Fig. 1

% Fig. 2

% Fig. 3

% Fig. 4

% Fig. 5

% Fig. 6

\newpage
% Fig. 7

\newpage
% Fig. 8

% Fig. 9

%%%%%%%%%%%%%%%%%%%%%%%%%%%%%%%%%%%%%%%%%


\begin{thebibliography}{10}

\bibitem{birmol97}
T.S. Bir\'o, D. Moln\'ar, Z. Feng, L.P. Csernai, {\it Phys. Rev.} {\bf D55}
  (1997) 6900.

\bibitem{csocse94}
T. Cs{\"o}rg{\H o} and L.P. Csernai, {\it Phys. Lett.} {\bf B333} (1994) 494.

\bibitem{csemis95}
L. P. Csernai and I.N. Mishustin, {\it Phys. Rev. Lett.} {\bf 74} (1995) 5005.

\bibitem{bjo83}
J.D. Bjorken, {\it Phys. Rev.} {\bf D27} (1983) 140.

\bibitem{LanLif2}
L.D. Landau and E.M. Lifshitz: {\it The Classical Theory of Fields} (Pergamon,
  1962).

\bibitem{Csernai94}
L.P. Csernai: {\it Introduction to relativistic heavy ion collisions} (Wiley,
  1994).

\bibitem{kapsri94}
J.I. Kapusta and A.M. Srivastava, {\it Phys. Rev. } {\bf D50} (1994) 5379.

\bibitem{gavmul94}
S. Gavin and B. M{\"u}ller, {\it Phys. Lett.} {\bf B329} (1994) 486.

\bibitem{fenmol97}
Z. Feng, D. Moln{\'a}r and L.P. Csernai, {\it Heavy Ion Phys.} {\bf 5} (1997)
  127.

\bibitem{rajwil93}
K. Rajagopal and F. Wilczek, {\it Nucl. Phys.} {\bf B399} (1993) 395.

\bibitem{rajwil93_2}
K. Rajagopal and F. Wilczek, {\it Nucl. Phys.} {\bf B404} (1993) 577.

\bibitem{gavgoc94}
S. Gavin, A. Gocksch and R.D. Pisarski, {\it Phys. Rev. Lett.} {\bf 72} (1994)
  2143.

\bibitem{asahua95}
M. Asakawa, Z. Huang, and X.N. Wang, {\it Phys. Rev. Lett} {\bf 74} (1995)
  3126.

\bibitem{huawan94}
Z. Huang and X.-N. Wang, {\it Phys. Rev.} {\bf D49} (1994) 4335.

\bibitem{S96}
L.P. Csernai, T.S. Bir{\'o}, Z.H. Feng, I.N. Mishustin, {\'A.} M{\'o}csy, D.
  Moln{\'a}r and O. Scavenius, {\it Heavy Ion Phys. } {\bf 4} (1996) 45.

\bibitem{birgre97}
T.S. Bir\'o and C. Greiner, {\it Phys. Rev. Lett.} {\bf 79} (1997) 3138.; C.
  Greiner and B. M{\"u}ller, {\it Phys. Rev.} {\bf D55} (1997) 1026.

\bibitem{NumRec}
W.H. Press, et al.: {\it Numerical recipes in C: the art of scientific
  computing} (Cambridge University Press, 1992 -- 2nd edition).

\bibitem{biaczy95}
A. Bialas, W. Czyz and M. Gmyrek, {\it Phys. Rev.} {\bf D51} (1995) 3739.

\bibitem{Huang66}
K. Huang: {\it Statistical Mechanics} (Wiley, 1966).

\bibitem{Reif65}
F. Reif: {\it Fundamentals of statistical and thermal physics} (McGraw-Hill,
  1965).

\bibitem{LanLif5}
L.D. Landau and E.M. Lifshitz: {\it Statistical Physics} (Pergamon, 1959).

\bibitem{blakrz96}
J.-P. Blaizot and A. Krzywicki, {\it Acta Phys. Pol.} {\bf B27} (1996) 1687.

\end{thebibliography}
\end{document}